\documentclass[10pt,conference]{IEEEtran}
\usepackage{xspace}

\usepackage{etoolbox}
\usepackage{times}
\usepackage{adjustbox}
\usepackage{caption}
\usepackage{subcaption}
\usepackage[super]{nth}
\usepackage{xcolor}
\usepackage{amsmath}
\usepackage{bbm}
\usepackage{flushend}
\usepackage{hyperref}
\usepackage[ruled,vlined]{algorithm2e}
\usepackage[normalem]{ulem}
\usepackage{booktabs}
\usepackage{wrapfig}
\usepackage{flushend}
\usepackage{enumitem}
\usepackage{multirow}
\usepackage{algorithmic}
\usepackage{fancyhdr}

\usepackage{cleveref}

\newcommand{\myparagraph}[1]{\noindent\textbf{#1. }}

\newcommand{\kbs}{Kubernetes}
\newcounter{mip}
\newcommand{\miplabel}[1]
{\refstepcounter{mip}\label{#1}MIP\textit{\themip}}

\newtoggle{showmarks}

\iftoggle{showmarks}{
  \newcommand\yanqi[1]{\textcolor{cyan}{YZ: #1}}
  \newcommand\sameh[1]{\textcolor{brown}{SE: #1}}
  \newcommand\delim[1]{\textcolor{magenta}{CD: #1}}
  
  \newcommand\TODO[1]{\textcolor{blue}{TODO: #1}}
}{
  \newcommand\inigo[1]{\unskip}
  \newcommand\gohar[1]{\unskip}
  \newcommand\rodrigo[1]{\unskip}
  \newcommand\ricardo[1]{\unskip}
  \newcommand\yanqi[1]{\unskip}
  \newcommand\sameh[1]{\unskip}
  \newcommand\delim[1]{\unskip}
  \newcommand\TODO[1]{\unskip}
  \newcommand\rf[1]{\unskip}
 }

\newcommand\camr[1]{\textcolor{black}{#1}}

\newcommand{\hpcayear}{2024}

\newcommand{\hpcasubmissionnumber}{NaN}
\title{Analytically-Driven Resource Management for Cloud-Native Microservices}

\def\hpcacameraready{} 

\newcommand\hpcaauthors{Yanqi Zhang$\dagger$, Zhuangzhuang Zhou$\dagger$, Sameh Elnikety$\ddagger$, Christina Delimitrou$\dagger\dagger$}
\newcommand\hpcaaffiliation{Cornell University$\dagger$, Microsoft Research$\ddagger$, MIT$\dagger\dagger$}
\newcommand\hpcaemail{yz2297@cornell.edu, zz586@cornell.edu, samehe@microsoft.com, delimitrou@csail.mit.edu}

\author{
  \ifdefined\hpcacameraready
    \IEEEauthorblockN{\hpcaauthors{}}
      \IEEEauthorblockA{
        \hpcaaffiliation{} \\
        \hpcaemail{}
      }
  \else
    \IEEEauthorblockN{\normalsize{HPCA \hpcayear{} Submission
      \textbf{\#\hpcasubmissionnumber{}}} \\
      \IEEEauthorblockA{
        Confidential Draft \\
        Do NOT Distribute!!
      }
    }
  \fi 
}

\fancypagestyle{camerareadyfirstpage}{%
  \fancyhead{}
  
  \fancyhead[C]{
    \ifdefined\aeopen
    \parbox[][12mm][t]{13.5cm}{\hpcayear{} IEEE International Symposium on High-Performance Computer Architecture (HPCA)}    
    \else
      \ifdefined\aereviewed
      \parbox[][12mm][t]{13.5cm}{\hpcayear{} IEEE International Symposium on High-Performance Computer Architecture (HPCA)}
      \else
      \ifdefined\aereproduced
      \parbox[][12mm][t]{13.5cm}{\hpcayear{} IEEE International Symposium on High-Performance Computer Architecture (HPCA)}
      \else
      \parbox[][0mm][t]{13.5cm}{\hpcayear{} IEEE International Symposium on High-Performance Computer Architecture (HPCA)}
    \fi 
    \fi 
    \fi 
    \ifdefined\aeopen 
      \includegraphics[width=12mm,height=12mm]{ae-badges/open-research-objects.pdf}
    \fi 
    \ifdefined\aereviewed
      \includegraphics[width=12mm,height=12mm]{ae-badges/research-objects-reviewed.pdf}
    \fi 
    \ifdefined\aereproduced
      \includegraphics[width=12mm,height=12mm]{ae-badges/results-reproduced.pdf}
    \fi
  }
  \fancyfoot[C]{}
}
\begin{document}
\maketitle

\ifdefined\hpcacameraready 
  \thispagestyle{camerareadyfirstpage}
  \pagestyle{empty}
\else
  \thispagestyle{plain}
  \pagestyle{plain}
\fi

\newcommand{\hpcaheight}{0mm}
\ifdefined\eaopen
\renewcommand{\hpcaheight}{12mm}
\fi


\begin{abstract}
Resource management for cloud-native microservices has attracted a lot of recent attention. 
Previous work has shown that machine learning (ML)-driven approaches outperform traditional techniques, such as autoscaling, in terms of both SLA maintenance and resource efficiency.
However, ML-driven approaches also face challenges including lengthy data collection processes and limited scalability. 
We present \textit{Ursa}, a lightweight resource management system for cloud-native microservices that addresses these challenges. Ursa uses an analytical model that decomposes the end-to-end SLA into per-service SLA, and maps per-service SLA to individual resource allocations per microservice tier. To speed up the exploration process and avoid prolonged SLA violations, Ursa explores each microservice individually, and swiftly stops exploration if latency exceeds its SLA. 

We evaluate Ursa on a set of representative and end-to-end microservice topologies, including a social network, media service and video processing pipeline, each consisting of multiple classes and priorities of requests with different SLAs, and compare it against two representative ML-driven systems, Sinan and Firm.
Compared to these ML-driven approaches, Ursa provides significant advantages: It shortens the data collection process by more than  $128\times$, and its control plane is $43\times$ faster than ML-driven approaches. At the same time, Ursa does not sacrifice resource efficiency or SLAs. During online deployment, Ursa reduces the SLA violation rate by $9.0\%$ up to $49.9\%$, and reduces CPU allocation by up to $86.2\%$ compared to ML-driven approaches.


\end{abstract}

\section{Introduction}
\label{sec:intro}

Production cloud services, such as Twitter and Netflix, are increasingly built as graphs of microservices~\cite{netflix_decompose, twitter_decomposing,gan2019open,Gan18b,Gan18}, and deployed with cloud-native frameworks like Kubernetes~\cite{k8s, aks, gke, ack}. Despite the benefits of modularity and elasticity, resource management for microservices that must meet SLA constraints, e.g., end-to-end latency, is challenging, due to the diverse resource requirement of individual microservices and their inter-service dependencies~\cite{gan2021sage, gan2019open}. 

Resource management for microservices has been the topic of recent studies, and machine learning (ML) models, especially deep neural networks (DNN), have become a popular choice to address the complexity of microservice topologies.
Previous studies have either used ML to predict important performance metrics, such as latency and load~\cite{zhang2021sinan, gan2021sage, chow2022deeprest, wang2022deepscaling, luo2022power}, or to directly adjust resource allocation~\cite{qiu2020firm}, and demonstrate that ML-driven approaches outperform traditional techniques, such as autoscaling~\cite{aws_step_scaling}, in performance and resource efficiency. 

However, ML-driven approaches still face key challenges limiting their adoption~\cite{Delimitrou13,Delimitrou13d,Delimitrou14,Delimitrou14b,Delimitrou15,Delimitrou16,Delimitrou17}.
First, ML-driven approaches typically require a lengthy exploration process to collect tens of thousands of data points to train the models, and therefore can not track changes in user behavior or handle the frequent updates to the microservice logic. 
Second, these ML models are on the critical path for every resource management decision, limiting the speed and scalability of resource management.
Third, previous studies are evaluated using conventional benchmarks that use remote procedure calls (RPC) as the only method of inter-service communication and include only lightweight text processing in the business logic~\cite{gan2019open, sriraman2018mu, zhou2018fault}. Modern cloud-native applications increasingly use both RPCs and message queues (MQ)~\cite{luo2021characterizing}, such as Kafka~\cite{kafka} and Redis streams~\cite{redis-streams}, handle different request classes, such as image processing and ML workloads~\cite{shahrad2020serverless, mahgoub2022orion, romero2021faast}, and support different request priorities.
Different request classes or priorities exhibit different latencies and therefore have different SLAs, making resource management more challenging. 

To address these challenges, we propose \textit{Ursa}, a lightweight resource management framework for cloud-native microservices. 
As a first step, we conduct a case study to understand how latency anomalies due to poor resource provisioning propagate through different communication methods. We show that backpressure is only significant for RPCs and is most pronounced in the parent service of the culprit tier (bottlenecked microservice). Given this, we design a method to determine the resource utilization threshold for each microservice that prevents backpressure in the application topology. 
By enforcing that the system operates in a backpressure-free zone, microservices in a topology can be treated as independent, reducing the number of factors the resource manager much consider from $O(N^2)$, where each pairwise microservice dependency must be accounted for, to $O(N)$, where the latency of individual microservices only depends on their own resource allocations. This greatly simplifies resource management, as most prior work resorts to complex ML models due to the need to capture the impact that microservice dependencies have on end-to-end performance. 

Moreover, in a backpressure-free system, we develop a performance model based on mixed integer programming (MIP) which decomposes end-to-end latency SLA constraints into per-service latency constraints, and maps them to resource allocation thresholds for individual microservices. 
The model also supports specifying different SLAs for different request classes and priorities.
To speed up the resource exploration, Ursa explores each microservice individually, and swiftly stops exploration when latency exceeds SLA or the resource utilization reaches its backpressure-free threshold. 

To better reflect modern microservices, we re-implemented several DeathstarBench applications~\cite{gan2019open} using Dapr~\cite{dapr}, a popular microservice framework developed and used by major cloud providers. The re-implemented benchmarks use both RPCs and message queues (MQs), and implement different request classes and priorities, executing more diverse business logic than before. 
We compare Ursa to two representative ML-driven systems, Sinan~\cite{zhang2021sinan} and Firm ~\cite{ qiu2020firm} as well as traditional autoscaling. Ursa reduces the required exploration time by more than $128\times$, making it more practical to track frequent changes to microservice logic. During online deployment, Ursa's control plane is $43\times$ faster than prior work, and Ursa reduces the SLA violation rate by $9.0\%$ to $49.9\%$, and the CPU allocation by up to $86.2\%$ compared to ML-driven approaches. 

\section{Related Work}
\label{sec:background}


\myparagraph{Microservices}
\camr{
Microservices have emerged as the dominant paradigm for interactive cloud services over the past few years. Unlike traditional monolithic architectures that contain the entire functionality of an application in a single binary, microservice architectures are graphical structures composed of tens or hundreds of single-purpose, loosely-coupled microservices, scaled independently, and in some cases implemented in different programming languages. The popularity of microservices is justified by several reasons, including flexible development, rapid iteration and fine-grained elasticity.
}
The emergence of microservices has also prompted efforts to benchmark and characterize them. Representative benchmarks include DeathstarBench~\cite{gan2019open} and ticket reservation~\cite{zhou2018fault}, which implement several end-to-end user-facing applications. 
These benchmarks use RPCs and \texttt{http} RESTful API as the only inter-service communication methods, and mostly perform lightweight text processing in the business logic. 
More recently, Luo et al.~\cite{luo2021characterizing} characterized microservices running on AliCloud and showed that MQs are common in practice, accounting for 23\% of all communication methods, and the performance of microservices is most sensitive to CPU interference. Related work~\cite{zhou2018overload, romero2021faast, mahgoub2022orion, shahrad2020serverless} also shows that cloud-native applications implement a variety of business logic, including ML workloads, webserving, image and video processing, etc.

\myparagraph{Resource management}
A large body of work has focused on using ML to adjust resource allocation for microservices. These systems use ML to predict important performance metrics, such as latency and load or diagnose performance issues~\cite{zhang2021sinan, gan2021sage, chow2022deeprest, wang2022deepscaling, luo2022power} or to directly adjust resource allocations~\cite{qiu2020firm,zhang2021sinan}. They demonstrate that ML-driven approaches outperform traditional approaches, such as autoscaling~\cite{aws_step_scaling} and queuing based mechanisms~\cite{powerchief}, in terms of performance and resource efficiency. 
\camr{However, ML-driven, especially deep learning driven approaches also suffer from demand of large training dataset, difficulties in adapting to changing application logic and user workload, and limited control plane scalability.}
In addition to using ML, Zhou et al.~\cite{zhou2018overload} reduce request failure rate in WeChat microservices with overload control, Yang et al.~\cite{powerchief} propose to identify bottleneck services in multi-phase applications by monitoring the queueing status, Suresh et al.~\cite{suresh2017distributed} adopt deadline-based scheduling to improve tail latency in multi-tier workloads, and Sriraman et al.~\cite{Sriraman18} present an auto-tuning framework for microservice concurrency, and show the impact of threading decisions on application performance and responsiveness.

Improving resource efficiency in cloud platforms in general is an important research area, and recent work~\cite{boutin2014apollo,ferguson2012jockey, ghodsi2011dominant,gog2016firmament, isard2009quincy,jyothi2016morpheus,park20183sigma, tumanov2016tetrisched,delgado2015hawk, karanasos2015mercury, ousterhout2013sparrow,Delimitrou14} has proposed several directions for how cluster scheduling frameworks can improve resource usage. Resource central~\cite{cortez2017resource} uses a set of ML models to predict VM performance metrics, such as CPU utilization and VM lifetime, Autopilot~\cite{rzadca2020autopilot} uses an ensemble of models to tune container configurations, Ambati et al.~\cite{ambati2020providing} propose providing SLOs for resource harvesting VMs, and Narayanan et al. ~\cite{narayanan2021solving} propose to efficiently solve large-scale resource allocation problems by partitioning them to smaller problems. However, these proposals are mainly applicable to single VMs or containers, rather than microservices with directed acyclic graph (DAG) topologies.


\section{Backpressure Effect}
\label{sec:backpressure}

\begin{figure*}[t]
\centering
\includegraphics[width=0.7\linewidth]{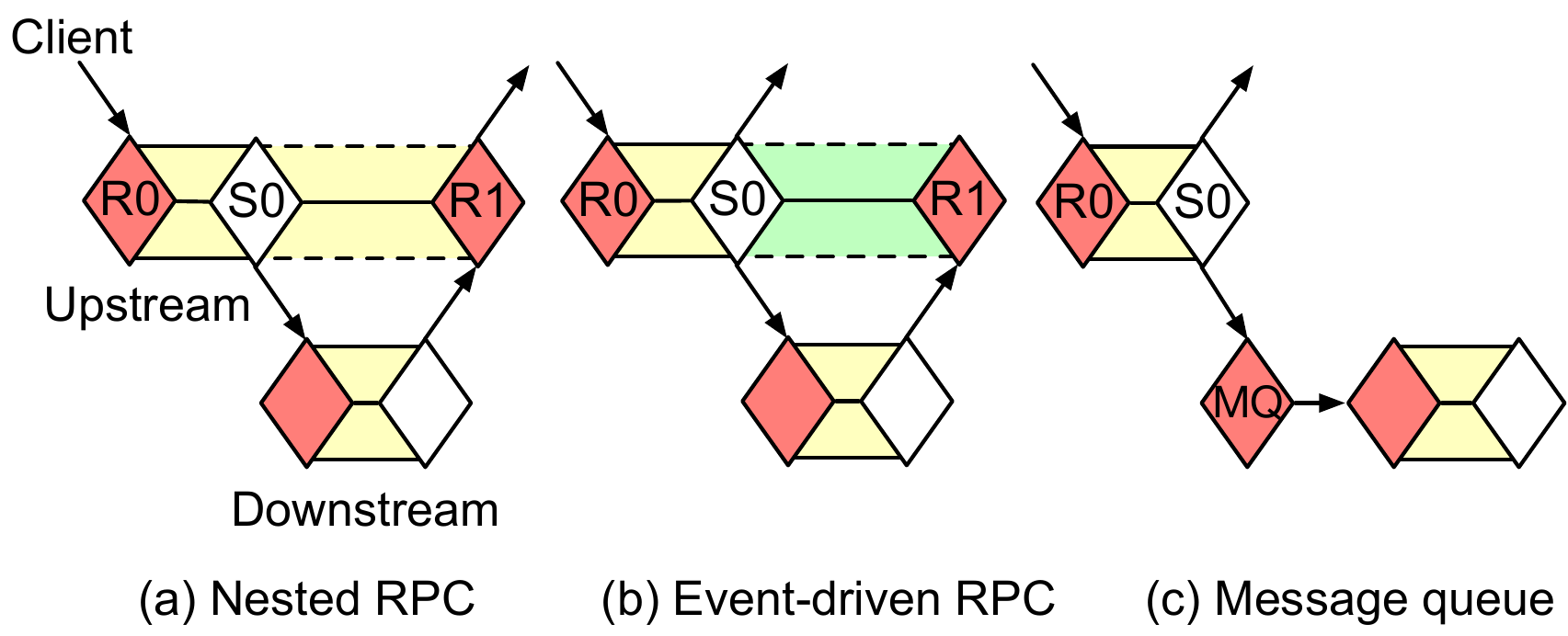}
\caption{Inter-service communication methods. }
\label{fig:commuication}
\vspace{-0.1in}
\end{figure*}

\begin{figure*}[t]
\centering
\hspace*{-0.1in}\includegraphics[width=0.78\linewidth]{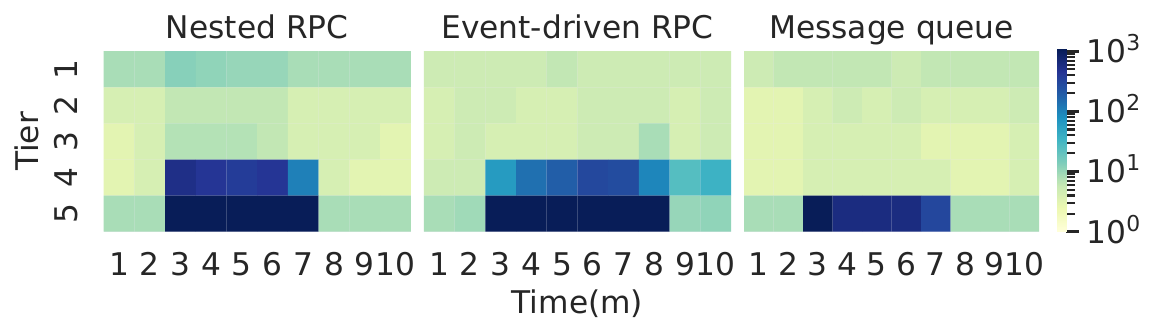}
\caption{Backpressure effects in a service chain. }
\label{fig:backpressure}
\vspace{-0.1in}
\end{figure*}

Backpressure is one of the major challenges in microservice resource management, and it refers to the phenomenon of the resource allocation of one service affecting the latency of upstream services, in addition to its own. 
In the presence of backpressure, modeling microservice latency requires modeling $O(N^2)$ dependencies, in the worst case, for a topology with $N$ microservices, as each microservice's latency can be affected by the resources of its downstream services.
Many microservice resource management frameworks use a centralized performance model that requires global information about all microservices to account for their inter-service dependencies, at the cost of scalability~\cite{zhang2021sinan, gan2021sage, chow2022deeprest}. 
To achieve scalable resource management for microservices, we first conduct a case study to understand how backpressure propagates through 
different communication methods, including RPCs and MQs. 

We study three types of chains connected by nested RPCs, event-driven RPCs, and message queues (MQs), respectively. 
Nested RPCs, as shown in Figure~\ref{fig:commuication}(a), are a synchronous system where, upon receiving the client request ($R0$), the upstream service forwards the request to the downstream service ($S0$) via RPC, blocks until the response is received ($R1$), and then returns the result to the client.
Event-driven RPCs~\cite{welsh2001seda}, as shown in Fig.~\ref{fig:commuication}(b), are more asynchronous in that upon receipt of a client request, the upstream service dispatches the request to another thread and returns immediately ($S0$), while the dispatched thread contacts the downstream service via RPC and waits for the response ($R1$).
\camr{It is noteworthy that event-driven RPCs are still not fully asynchronous, as they involve a two-step process. Upon receiving a user request, the event handling thread works asynchronously, by yielding immediately after creating a daemon thread that further processes the request. However, the daemon thread talks to the downstream service synchronously and informs the event thread when it receives the response. This interaction involves maintaining synchronous connections between upstream and downstream services.}
Message queues, on the other hand, are completely asynchronous. Unlike RPCs, MQs, such as Kafka~\cite{kafka} and Redis streams~\cite{redis-streams}, 
mostly use a publish-subscribe paradigm, where publishers publish messages to topics hosted by the MQ and subscribers consume messages by subscribing to the topics. As shown in Figure~\ref{fig:commuication}(c), the upstream service does not directly contact the downstream service, but instead sends the client request to the MQ, and the downstream service gets new requests by polling the MQ. 

\myparagraph{Characterizing backpressure} 
We implement the RPC service chains with gRPC~\cite{gRPC} and the MQ with Redis streams~\cite{redis-streams}. Each chain is configured to include 5 tiers, with each tier implementing a CPU-intensive loop as the request handler. We record the per-tier \emph{response time} ($S0-R0$), which in the cases of MQ is the service latency, and in the case of RPC is the service latency \textit{excluding the duration waiting for the downstream response}. We measure the tier's response time because it is closely related to the resource allocation of the tier itself. 
We stress test each service chain for 10 minutes, injecting performance anomalies into the leaf tier (tier 5) by throttling its CPU limit between minutes 3 and 6.
The resulting backpressure behavior is shown in Figure~\ref{fig:backpressure}, where each column on the x-axis represents a one minute interval, each row on the y-axis corresponds to a tier (tier 1 is client-facing), and the color of each cell highlights the per-tier $99^{th}$ response time during that minute. 
For both nested and event-driven RPCs, significant backpressure is observed, especially for tier 4, the parent of the throttled leaf tier, 
and the backpressure rapidly diminishes up the call chain and becomes negligible above tier 3. 
In contrast, MQ shows no backpressure behavior, even on tier 4.


\begin{figure}[t]
\centering
\includegraphics[width=.95\linewidth]{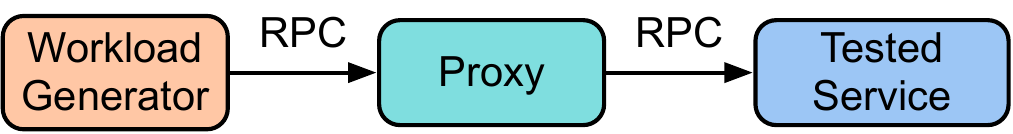}
\caption{Backpressure profiling engine architecture.}
\label{fig:offline_arch}
\vspace{-0.2in}
\end{figure}

\begin{figure*}[t]
\centering
\includegraphics[width=.78\linewidth]{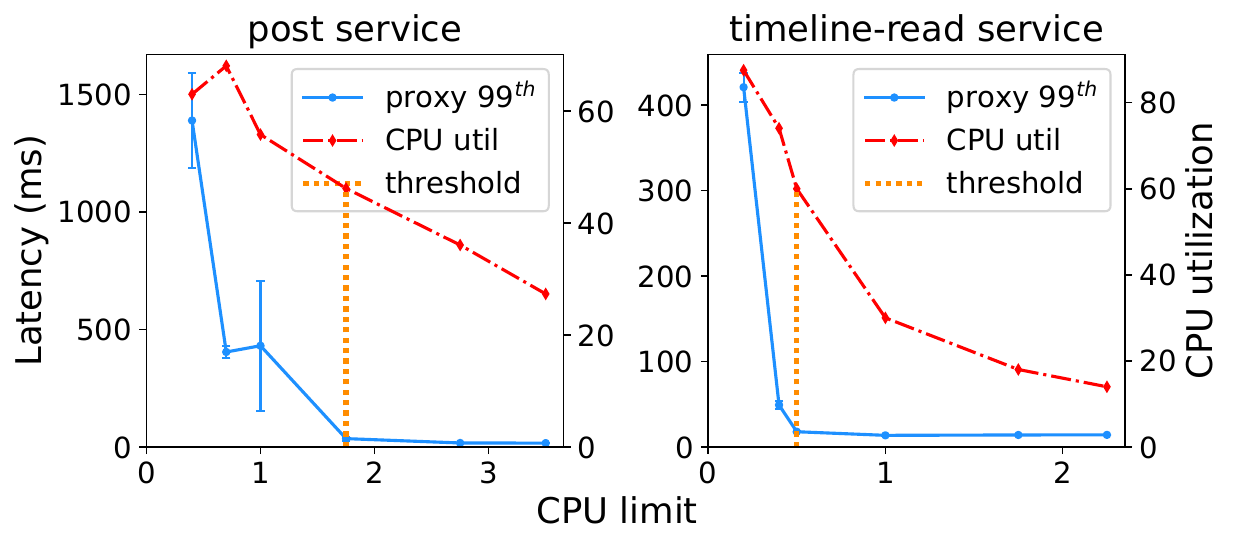}
\caption{Identifying backpressure-free CPU thresholds in a service mesh. We incrementally decrease the amount of resources allocated to the tested microservice, until we observe an increase in the latency of the proxy. Given the proxy's lack of computation activity, this increase signals the presence of backpressure. We use this threshold as the utilization the tested service should not exceed to avoid introducing backpressure to its parent tiers. }
\label{fig:socialnet_backpressure}
\end{figure*}

\vspace{0.08in}
\myparagraph{Determining conditions for negligible backpressure} 
Backpressure complicates resource management because the latency of a service also depends on the resources of its downstream services, in addition to its own. To simplify resource management, a natural approach is to determine safe CPU utilization thresholds to avoid backpressure in the system, as the performance of microservices is most sensitive to CPU utilization~\cite{luo2021characterizing, gan2019open}. The approach generalizes to other resources as well, for microservices with different resource profiles. 
To this end, we use a profiling engine with the 3-tier architecture shown in Figure~\ref{fig:offline_arch}, where the proxy acts as the parent service and simply forwards the request to the tested service via RPC.
The engine gradually increases the CPU limit of the tested service, and monitors the latency of the proxy and the CPU utilization of the tested service, until the latency of the proxy converges. The convergence of proxy latency is determined by comparing the latency recorded under the last two CPU limits with Welch's t-test~\cite{welch1947generalization}, a classical hypothesis testing method for identifying whether the means of the two sets of samples are equal.
The CPU utilization just before the convergence of the proxy latency is then recorded as the threshold for not triggering backpressure. 
\camr{
In order to account for complex invocation patterns, such as multiple upstream services sending requests to a downstream service concurrently, the workload generator synthesizes aggregate loads from different upstream services, so that the measured backpressure-free threshold is valid under fan-in/out patterns.
}

Figure~\ref{fig:socialnet_backpressure} shows the profiling process for two microservices in a social network application similar to~\cite{gan2019open}: the post service, responsible for querying the contents of user post, and the timeline-read service, responsible for querying the post IDs in user timelines.
The x-axis corresponds to the CPU limit of the tested service. The left and right y-axis correspond to latency and CPU utilization, respectively.
The blue and green lines show the average $99^{th}$ percentile latency of the proxy and the tested service under different CPU limits, and the error bars represent the standard deviation. The red line indicates the CPU utilization of the tested service. 
The orange line highlights the point where the proxy latency converges, and the corresponding CPU utilization is recorded as the threshold, $46.2\%$ for post service and $60.0\%$ for timeline-read service. 
When significant backpressure is observed, the $99^{th}$ percentile latencies of the proxy and the tested service have already increased by more than $5\times$ and $10\times$.

\vspace{0.08in}
\myparagraph{Main insights} The study produces the following insights.
\begin{enumerate}[leftmargin=1.75\parindent]
    \item Backpressure complicates resource management because the latency of a service depends on resources of its downstream services, in addition to its own. Backpressure is common in RPCs but negligible in MQs. 
    \item Backpressure diminishes along the invocation chain. Of all the upstream services of the culprit,
    the parent service shows the most significant increase in latency. 
    \item The backpressure-free resource utilization threshold of a service can be profiled by monitoring the latency of an upstream proxy. By operating within the thresholds, backpressure can be avoided in the microservice system. 
    \item By eliminating backpressure, the number of factors required to model microservice latency becomes $O(N)$ with $N$ services, because a microservice's latency is only a function of its own resources. Otherwise the number is $O(N^2)$ in the worst case, as a microservice's latency may depend on resources of its downstream services.
\end{enumerate}
\section{Performance model}
\label{sec:performance_model}

When the system has negligible backpressure, the latency distribution of a microservice becomes mainly a function of its own resources. 
We now build a performance model for mapping microservice SLAs to resources. First, we decompose the end-to-end latency constraint for each request type to a set of per-microservice latency constraints. Second, we map each per-microservice latency constraints to resources for that service. 
In this section we describe the performance model we use, based on mixed-integer programming. 

\vspace{0.08in}
\myparagraph{Decomposing end-to-end latency} Without loss of generality, we consider the end-to-end latency of a chain that handles a single type of requests, which is the basic structure of microservice DAGs and to which other topologies can be transformed. For examples, a tree consists of multiple chains from the root to the leaf services, and similarly, fan-in and fan-out topologies can be decomposed to multiple chains from the source to the sink services. If a service is accessed multiple times by an upstream service, we consider the cumulative latency of all accesses as the latency of that service. 

\textbf{Theorem 1:} Consider a chain of microservices $S_1$ to $S_n$, and their response time distributions $t_1$ to $t_n$, where $t_i(x_i)$ is the $x_i^{th}$ percentile latency of microservice $S_i$, $x_i \in [0, 100]$. Similarly, we define $t_e$ to be the end-to-end latency distribution, and $t_e(x_e)$ to be the $x_e^{th}$ percentile end-to-end latency, $x_e \in [0, 100]$. Then,
\begin{align*}
    t_e(x_e) \leq \sum_{i=1}^{n} t_i(x_i),~\mbox{if}~100-x_e \geq \sum_{i=1}^n 100-x_i \stepcounter{equation}\tag{\theequation}\label{thoerem}
\end{align*}

The theorem holds true regardless of the joint distribution of microservice latencies (i.e., if services are independent or correlated), and it denotes that \textit{the sum of per-microservice latencies provides an upper bound for the end-to-end latency at an arbitrary percentile, as long the sum of residuals of per-microservice percentiles is no greater than the residual of the end-to-end percentile.}
The proof can be found in the supplementary material. 

Theorem 1 proposes a method to \emph{guarantee the end-to-end latency SLA by examining the latency of individual microservices}.
For example, in a chain consisting of two microservices $S_1$ and $S_2$, with the SLA defined as the end-to-end $99^{th}$ percentile latency, Theorem 1 suggests that the actual end-to-end $99^{th}$ percentile is less than the sum of $x_1^{th}$ percentile latency of $S_1$ and $x_2^{th}$ percentile latency of $S_2$, as long as $100-x_1+100-x_2 \leq 1$, and in other words, $(x_1, x_2)$ can be $(99.1, 99.9)$, $(99.5, 99.5)$, $(99.7, 99.3)$, etc. 
Since all such combinations of $x_i$ are upper bounds of the actual end-to-end latency, the end-to-end SLA must be satisfied as long as the corresponding sum of the per microservice latencies for one combination is less than the end-to-end SLA target. 
More generally, given the end-to-end latency SLA of $x_e^{th}$ percentile latency needing to be less than $T$ in a chain of length $n$, the end-to-end SLA is satisfied if
\begin{align*}
    \exists [x_1...x_n] ~s.t.~ \sum_{i=1}^{n} t_i(x_i) \leq T~\&~100-x_e \geq \sum_{i=1}^n 100-x_i \stepcounter{equation}\tag{\theequation}\label{eq:constraint}
\end{align*}

\vspace{0.08in}
\myparagraph{Mapping per-microservice latency to resources} 
To optimize resource allocation, the per-microservice latency distributions $t_i$ need to be associated with resources, to derive a model that maps SLAs to resources. 
In addition, the model should be able to handle multiple classes or priorities of requests instead of a single class as in Theorem 1.

In cloud-native frameworks, such as Kubernetes~\cite{k8s, aks, gke}, dynamic resource tuning is typically achieved by the changing the number of replicas, \camr{or container instances}, each with a predefined resource configuration (CPU and memory). Therefore, we use \emph{load per replica (LPR)} as the metric to relate resources to latency, where load is measured in requests per second (RPS). 
Considering a service $S_i$ that handles $c$ classes or priorities of requests ($v_1$ to $v_c$), the load per replica $y_i$ can be represented as a vector $[a_i^1 ... a_i^c]$, where $a_i^c$ is the load for request class $v_c$. If the load per replica vector $y_i$ is used as the resource allocation threshold and the total load to $S_i$ is $[A_i^1 ... A_i^c]$, the resources consumed by $S_i$ can be calculated using Equation~\ref{eq:resource}, where $u_i$ is the resource consumption per replica. 

\begin{align*}
    r_i(y_i) = \max_{1 \leq j \leq c}\lceil \frac{A_i^j}{a_i^j} \rceil \cdot u_i \stepcounter{equation}\tag{\theequation}\label{eq:resource}
\end{align*}


On the other hand, since the latency distribution of request $v_j$ in $S_i$ is a function of LPR $y_i$, the $x_i^{th}$ percentile latency of $v_j$ can be denoted as $t_i^j(y_i, x_i)$. Then
the $t_i(x_i)$ items in Equation~\ref{eq:constraint} can be replaced by $t_i^j(y_i, x_i)$, transforming the latency constraint to a resource allocation constraint.
While $t_i^j(y_i, x_i)$ can be fitted with profiling data, the resource-latency function can be an arbitrary non-increasing function that is not necessarily convex, which makes it hard to use convex optimization models. Instead, we can discretize the variables and use the function in Mixed Integer Programming (MIP), which can be efficiently solved by modern optimization solvers using heuristics, such as branch-and-bound algorithm~\cite{land2010automatic}.
Specifically, we discretize the percentile variable $x_i$ and LPR $y_i$, and represent the latency distributions under different LPRs as a matrix $D_i^j$, where each element of $D_i^j$ is the latency that corresponds to a certain percentile for a given LPR. For example, assuming that $S_i$ is profiled under $m$ different LPRs $Y_i=[y_i^1...y_i^m]$ and the latency distribution is discretized into $h$ different predefined percentiles $P=[p_1...p_h]$, $D_i^j$ will be a $m \times h$ matrix where $D_i^j[\alpha, \beta]$ is the latency at percentile $p_{\beta}$ under LPR $y_i^{\alpha}$.
As a result of the discretization, the LPR variable $y_i$ can be represented by a one-hot vector $\delta_i$ of length $m$, indicating which LPR is chosen as the resource allocation threshold. Similarly, the percentile variable can be presented by one-hot vector $\gamma_i^j$ of length $h$, indicating which percentile contributes to thesum of per-microservice latencies for request class or priority $v_j$. 
With the two one-hot decision variables, the latency of request class or priority $v_j$ in microservice $S_i$ can be expressed as $\delta_i^T D_i^j \gamma_i^j$, and the resource consumption can be expressed as $\delta_i^T R_i$, in which $R_i$ is a 1-D vector corresponding to resource consumption under the profiled LPRs, computed with Equation~\ref{eq:resource}. 

\vspace{0.08in}
\myparagraph{Resource optimization model} 
Given that we can provide an upper bound on the end-to-end latency using the sum of per-microservice latencies and map per-microservice latencies to resource allocation thresholds, we can now design an optimization model that calculates the most efficient resource allocation threshold for each microservice, such that they all meet their respective per-microservice SLAs. Specifically, the inputs to the model include the load of the application, SLAs for different request classes and priorities, and per-microservice latency distributions under different LPR thresholds. The output of the model is the most efficient per-microservice LPR threshold that satisfies SLAs.
With the described notations summarized in Table~\ref{tab:mip_notations}, we derive the following solvable mixed-integer programming (MIP) model that yields optimal resource configurations given a set of end-to-end constraints: for each request class or priority $v_j$, the $x_j^{th}$ percentile latency should be less than $T_j$.

\begin{table}[h!]
\begin{normalsize}

\begin{tabular}{ll}
\toprule
\textbf{}   & \textbf{Description}                               \\ \midrule
$\delta_i$           & Resource (LPR) one-hot vector             \\ 
$\gamma_i^j$         & Latency percentile one-hot vector                 \\
$R_i$                & Resource consumption under different LPRs \\
$P$                  & Discretized percentile values             \\
$D_i^j$              & Latency distribution matrix               \\
$T_j$                & End-to-end SLA target               \\
$x_j$                & End-to-end SLA target percentile          \\
$\mathbbm{1}$        & 1-D vector whose elements are all 1       \\
\bottomrule
\end{tabular}
\end{normalsize}

\caption{\label{tab:mip_notations} Notations in the MIP formulation.}
\end{table}

\vspace{-0.2in}
\begin{quote}
    \[
        \begin{array}{llll} 
            \mbox{minimize}& \sum_{i=1}^{n} \delta_i^T R_i,  \\
            \mbox{subject to}& \sum_i \delta_i^T D_i^j \gamma_i^j \leq T_j, \forall j \hspace{5em} (1) \\
                             & \sum_i 100 - P^T \gamma_i^j \leq 100 - x_j, \forall j \hspace{0.5em}(2) \\
                             & \mathbbm{1}^T \delta_i = 1, \forall i \hspace{8.5em}(3) \\
                             & \mathbbm{1}^T \gamma_i^j = 1, \forall i, j \hspace{7.35em}(4) \\
            \mbox{variables}&~ \delta_i~( 0 \leq \delta_i \leq 1~\&~ \delta_i \in Z) \\
                            &~ \gamma_i^j~( 0 \leq \gamma_i^j \leq 1~\&~ \gamma_i^j \in Z)
        \end{array}
        \miplabel{mip:rsc}
    \]
\end{quote}

The objective of MIP~\ref{mip:rsc} is to minimize the total resource consumption, while meeting the end-to-end SLA. Constraint 1 specifies that for each request class or priority, the sum of per-microservice latencies must be smaller than the SLA target, and constraint 2 specifies that the sum of per-microservice latencies in the constraint 1 is an upper bound of the actual end-to-end latency. The rest of the constraints enforce the decision variables to be one-hot vectors. 
For each microservice, the LPR one-hot vector $\delta_i$ produced by MIP~\ref{mip:rsc} corresponds to the most efficient resource allocation threshold among all profiled LPRs, which allows resource allocation of each microservice to be decided independently, by simply checking the load of the microservice. 

\vspace{0.08in}
\myparagraph{Mitigating latency overestimation} The quality of the solution of MIP~\ref{mip:rsc} is related to the tightness of the upper bound given by Theorem 1, as a loose upper bound well above the actual latency can lead to overprovisioning of resources. An intuitive way to tighten the upper bound is to record the ratio of the upper bound to the actual value and use that ratio to refine the SLA constraint in MIP~\ref{mip:rsc}. For example, if the overestimation ratio of request class or priority $v_j$ is $\alpha_j$ and its expectation is $E(\alpha_j)$, constraint 1 in MIP~\ref{mip:rsc} can be refined to $\sum_i \delta_i^T D_i^j \gamma_i^j \leq E(\alpha_j) T_j$.
With a fixed resource allocation denoted by $\delta_i^*$, for microservice $S_i, \forall i$, the upper bound on the latency of request class or priority $v_j$ can be solved using MIP~\ref{mip:lat}. The objective value is the tightest upper bound, because any percentile combination satisfying constraint 1 in MIP~\ref{mip:lat} establishes a upper bound on latency, and the objective is the smallest among all these upper bounds. Thus, the overestimation ratio $\alpha_j$ is the ratio of the objective over the actual latency, and $E(\alpha_j)$ is the average of $\alpha_j$ with different resource allocations.

\begin{quote}
    \[
        \begin{array}{lll} 
            \mbox{minimize}& \sum_i \delta_i^{*T} D_i^j \gamma_i^j,  \\
            \mbox{subject to}&  \sum_i 100 - P^T \gamma_i^j \leq 100 - x_j, \forall j \hspace{1em}(1) \\
                             & \mathbbm{1}^T \gamma_i^j = 1, \forall i, j \hspace{7.85em}(2) \\
            \mbox{variables}&~ \gamma_i^j~( 0 \leq \gamma_i^j \leq 1~\&~ \gamma_i^j \in Z)
        \end{array}
        \miplabel{mip:lat}
    \]
\end{quote}

\camr{In practice, instead of using the expected overestimation ratio to tighten the bound, one can choose other metrics, such as the overestimation ratio at a high percentile, to tradeoff the potential resource efficiency and the risk of SLA violation. We leave this exploration to future work. }


\vspace{0.08in}
\myparagraph{Discussion} In Ursa, we use this performance model to find the most efficient resource allocation given per-microservice latency SLA constraints, however, the model can be extended to other cases with minor modifications. For example, the model can handle end-to-end latency minimization under resource constraints, by replacing the objective of MIP~\ref{mip:rsc} with the sum of per-microservice latencies, and using the total available resources as a constraint. 

In addition, the model can handle SLAs defined in terms of request failure rates. The failure rate of an end-to-end request is no greater than the sum of request failure rates of the microservices it goes through, and each microservice's request failure rate is related to its resources, since insufficient resources will cause requests to time out and fail. The sum of per-microservice failure rates can then be used as a constraint to strengthen MIP~\ref{mip:rsc}.
The model can also support dynamic request paths by adding recorded paths to the model during deployment. In the case of a microservice being accessed multiple times in a dynamic path, the model can be simplified by considering the total time spent in each microservice.
We plan to investigate these potential use cases in future work. 

\begin{algorithm}[!ht]
\SetAlgoLined
 \textbf{Input:} Initial replica $R$, SLA violation threshold $F_{sla}$, backpressure-free threshold $CPU_{bp}$, profiling time $T$\;
 \textbf{Output:} Mapping from $LPR$ to latency distributions\;
 \textbf{Variable: } Replica $r$, replica tuning step $step$, Load~$L$, SLA violation frequency $f_{sla}$, CPU utilization $cpu$, latency distribution $d_{lat}$\;
 Initialize $r \leftarrow R$,~$map \leftarrow \{\}$\;
 \While{$r > 0$}{
  $wait(T)$\;
  \uIf{$cpu \geq CPU_{bp}$ || $f_{sla} \geq F_{sla}$} {
    \textbf{terminate}\;
  } \Else {
    $map[\frac{L}{r}] = d_{lat}$, $r = r - step$\;
  }
 }
 \Return $map$
 \caption{LPR threshold profiling algorithm. }
 \label{algo:service_exploration}
\end{algorithm}

\begin{figure*}[t]
\centering
\includegraphics[width=0.66\linewidth]{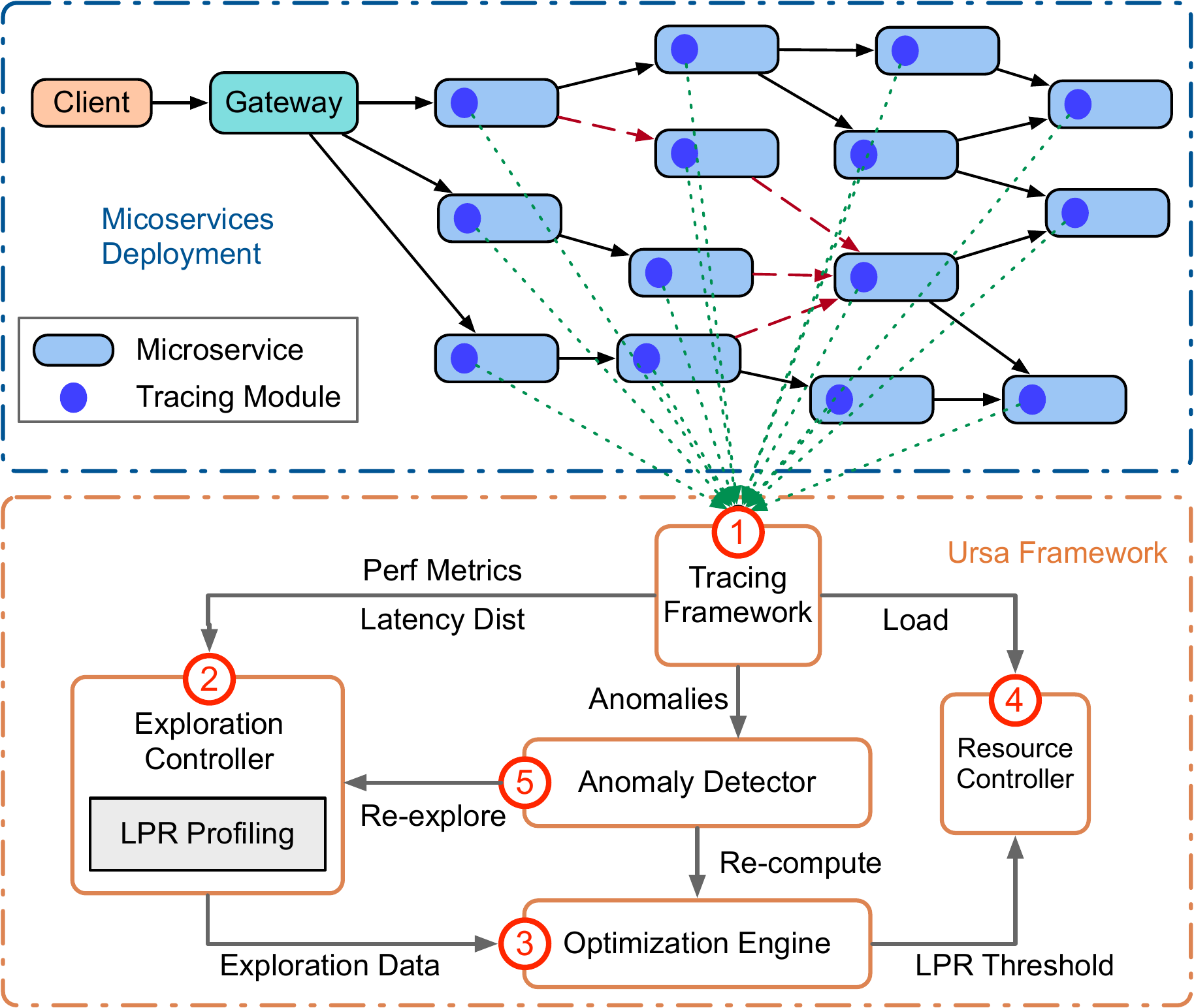}
\caption{System architecture of Ursa. }
\label{fig:design}
\vspace{-0.2in}
\end{figure*}

\vspace{0.08in}
\myparagraph{Allocation space exploration} The task of allocation space exploration is to collect input data for the MIP model, including the potential resource allocation thresholds and the corresponding latency distributions for each microservice. 
Exploration should include the most efficient resource allocation threshold for each microservice and converge fast, by terminating swiftly if the latency exceeds the SLA. To achieve this goal, we explore each microservice individually with the LPR threshold profiling algorithm shown in Algorithm~\ref{algo:service_exploration}, by replaying the workload trace on the profiled microservice. 
During profiling, we gradually reduce the replicas of the profiled microservice to increase the load on each replica and record the corresponding latency distributions. Profiling is terminated when SLA violations are observed. Additionally, profiling is also terminated when the CPU utilization of the microservice exceeds the service's backpressure-free threshold to preserve the independence assumption used by the performance model.

\section{Design and Implementation}
\label{sec:design}

We now present the design and implementation of \textbf{\textit{Ursa}}, a resource management framework based on the proposed performance model and allocation space exploration mechanism. 
Ursa is built on top of \kbs~\cite{k8s}, a popular container orchestration framework adopted by major cloud providers~\cite{eks, aks, gke, ack}, and leverages \kbs's APIs to dynamically allocate resources by tuning the number of replicas per microservice. 
Ursa requires the user to provide the topology and the end-to-end SLAs of the microservice application, including request paths, percentiles, and target latencies. 

Ursa aims to make allocation decisions fast and scalable. Ursa simplifies resource management decisions to threshold-based scaling by implementing the performance model from Section~\ref{sec:performance_model}. In addition, Ursa collects input data for the performance model via the exploration process used to identify the relation between resources and per-microservice SLAs. The components of Ursa are shown in Figure~\ref{fig:design}, and the functionality of each component is described below.


1. The \textbf{tracing framework} is implemented with Prometh-eus~\cite{prometheus}, a time-series database for metrics monitoring. It collects the CPU and memory usage data, as well as the request counts and latency distributions of each service which are used to calculated statistics required by the MIP model.

2. The \textbf{exploration controller} implements the allocation space exploration mechanism. It first determines the backpressure free CPU utilization thresholds for each RPC-connected microservice individually (Section~\ref{sec:backpressure}). Then, it explores feasible resource allocation thresholds of each microservice individually using Algorithm~\ref{algo:service_exploration}. \camr{The complete exploration of all microservices is required for a new application and must be completed before Ursa begins to manage the application. In the case of continuous development of individual microservices, only the updated microservices require re-exploration.}

3. The \textbf{optimization engine} then determines the resource allocation threshold for each microservice using the performance model in Section~\ref{sec:performance_model}, using exploration data and user load information collected by the tracing framework. The optimization engine is implemented with Gurobi~\cite{gurobi}.
\camr{During the initial deployment phase, the optimization engine is invoked with current user load information to generate the optimal scaling thresholds of each microservice. Subsequently, the optimization engine may need to be invoked again if the mix of user requests changes significantly or when the business logic of microservices is updated.}


4. Using the load per replica (LPR) threshold calculated by the optimization engine, the \textbf{resource controller} dynamically adjusts the number of replicas as the load changes, ensuring that for any class or priority of request, the average load on each replica does not exceed the threshold. 
Specifically, the resource controller determines whether the average load in one replica exceeds the threshold using Welch’s t-test~\cite{welch1947generalization} to accommodate the noise of load fluctuations. \camr{Specifically, the resource controller compares the actual load of the microservice, with the recorded load used as the scaling threshold, and considers the threshold to be exceeded if the t-test rejects the hypothesis that the mean of the actual load is less than the mean of the recorded load.}

5. During deployment, the \textbf{anomaly detector} periodically checks for anomalies in load and latency, and triggers recalculation of resource allocation thresholds or re-exploration, if necessary. 
Load anomalies refer to drastic changes in the ratio of different classes or priorities of requests that may lead to resource over-provisioning, in which case resource allocation thresholds are recalculated to improve resource efficiency. 
The anomaly detector identifies changes in request ratios by monitoring the \textit{request ratio deviation} of each microservice, which measures the difference between the load of the microservice and the load per replica threshold for scheduling. The metric is denoted by $\max_{i} \frac{l_i}{t_i} \frac{\sum_{i} t_i}{\sum_{i} l_i}$, where $l_i$ and $t_i$ are the total load and per-replica load threshold for the $i^{th}$ request class or priority. 
When the request ratio deviation exceeds a user-defined threshold,
the anomaly detector asks the optimization engine to recalculate the thresholds and update the resource controllers. 
If the re-calculated thresholds still fail to mitigate the request ratio deviation, indicating that the load pattern is not covered by previous exploration, the anomaly detector asks the exploration controller to re-explore the affected microservice. 

On the other hand, latency anomalies refer to SLA violations, which indicate that the latency distribution recorded during exploration needs to be updated. Similar to load anomalies, users can specify an end-to-end SLA violation threshold that triggers the re-exploration process if the SLA violation exceeds the threshold during deployment.


\section{Microservice Benchmarks}
\label{sec:benchmark}



Conventional microservice benchmarks~\cite{gan2019open, sriraman2018mu, zhou2018fault} have several limitations. 
First, conventional benchmarks use RPCs as the only method for inter-service communication, whereas MQs are becoming increasingly common in practice~\cite{luo2021characterizing}. 
Second, the business logic of conventional benchmarks involves only lightweight text processing, whereas a modern microservice handles different user request classes performing tasks, such as image processing and ML workloads~\cite{shahrad2020serverless, mahgoub2022orion, romero2021faast}, and even different request priorities, making resource management more challenging.
To address these limitations, we implement three benchmark applications using Dapr~\cite{dapr}, a popular microservice framework developed and used by major cloud providers, as described below. 
For all the applications, we implement the business logic in Golang and Python, and use gRPC~\cite{gRPC} for RPCs, Redis streams~\cite{redis-streams} for message queues and Redis~\cite{redis} for data stores. We incorporate MQ into our approach because it is an increasingly popular communication framework for microservices. 

\camr{
In our benchmarks, the interactive functionalities demanding immediate response, such as reading timelines from the social network, are implemented with RPC. Conversely, functionalities that do not require immediate response, such as labeling objects in user-uploaded images, are implemented with message queues.
}

 \begin{figure}[t]
 \centering
 \includegraphics[width=.95\linewidth]{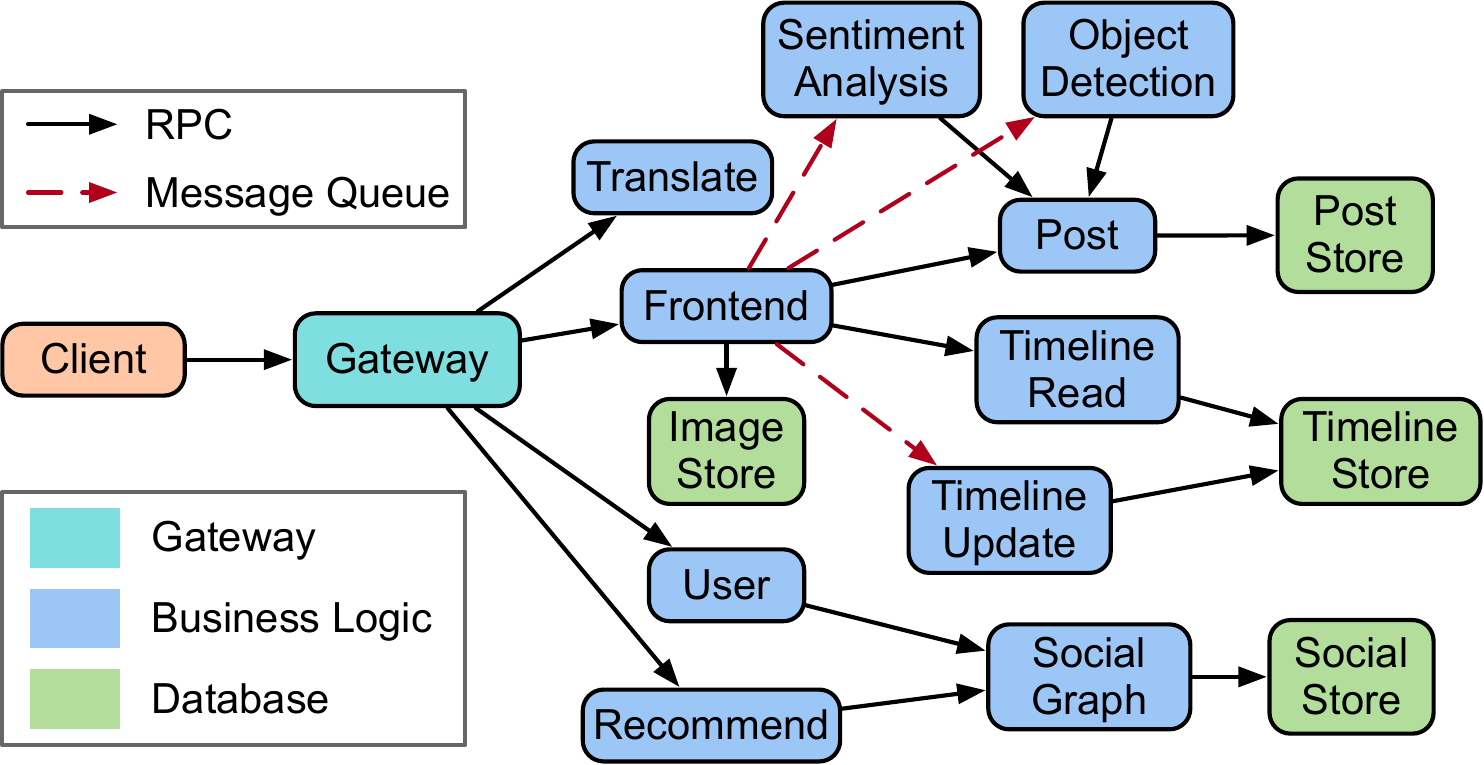}
 \caption{Social network microservice topology.}
 \label{fig:socialnet}
 \end{figure}

\vspace{0.05in}
\myparagraph{Social network} The social network application is a re-implementation of the DeathStarBench~\cite{gan2019open} application. 
In addition to the original features including uploading text posts and reading timelines, the re-implemented version includes several new features, including uploading images, sentiment analysis of texts, and object detection of images. Sentiment analysis and object detection are implemented with machine learning models from Hugging Face~\cite{huggingface}, and are connected to other services via MQs. 

 \begin{figure}[t]
 \centering
 \includegraphics[width=.95\linewidth]{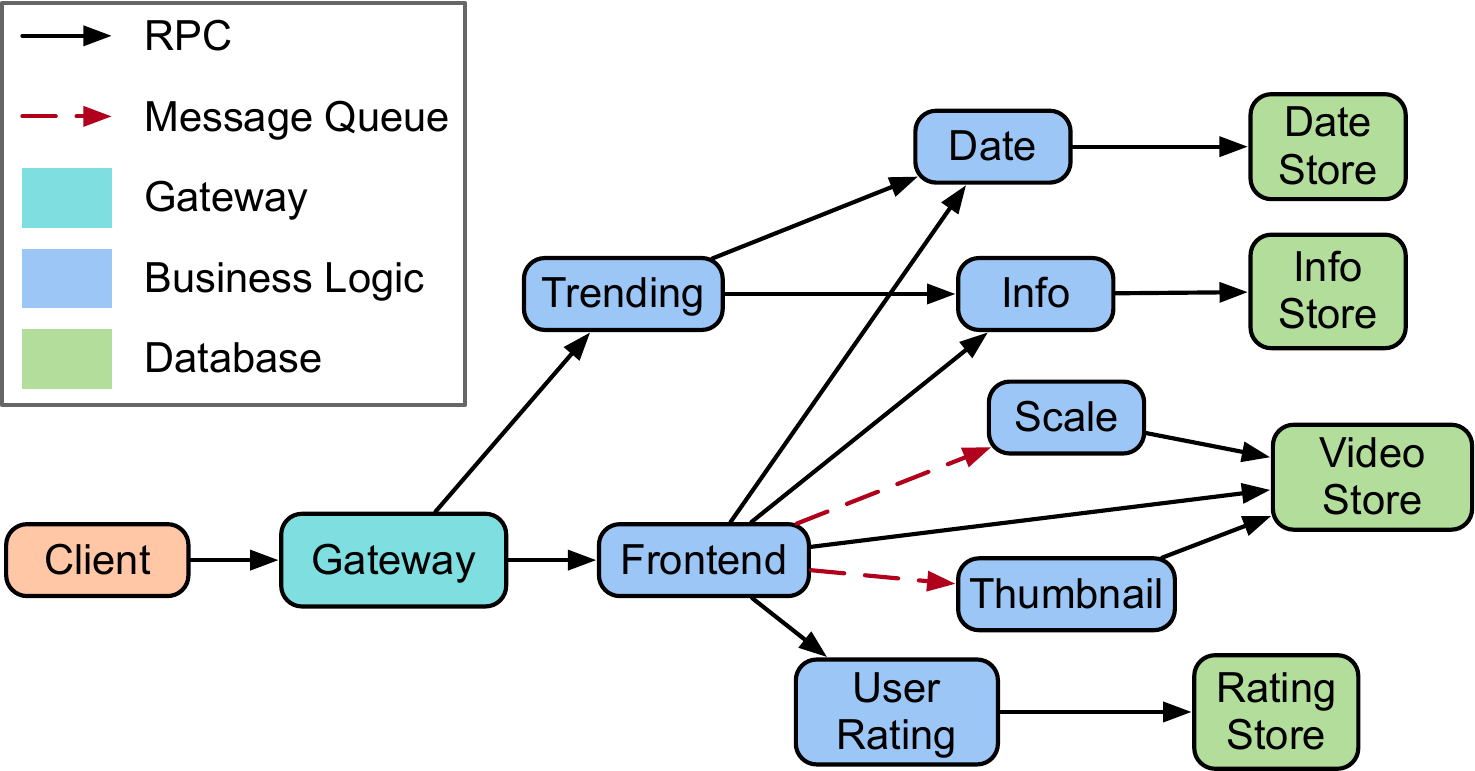}
 \caption{Media service topology.}
 \label{fig:video-sharing}
 \end{figure}

\vspace{0.05in}
\myparagraph{Media service} The media service is also a re-implementation of the corresponding DeathStarBench application. In addition to the original features including reviewing and rating videos, the re-implemented version additionally allows users to upload and download actual videos, and includes video-processing tasks, such as transcoding to different resolutions and generating thumbnails via FFmpeg~\cite{ffmpeg}. The video transcoding and thumbnail services are connected to other services via MQs.

\vspace{0.05in}
\myparagraph{Video processing pipeline} Video processing pipeline consists of three stages: The first stage extracts video metadata, the second stage takes snapshots from the video at fixed intervals, and the third stage performs face recognition on the video snapshots. The first two stages use FFmpeg, the third stage uses OpenCV~\cite{opencv}, and stages are connected with MQs.
The application handles two request priorities. High-priority requests are always processed immediately when worker threads are  available, while low-priority requests are processed only when there is no high-priority request waiting in the queue. 

Previous work typically handles a single SLA and only manages synchronous requests. For example, Sinan~\cite{zhang2021sinan} handles a single SLA of 500ms for the $99^{th}$ percentile latency of upload-post, read-timeline, and update-timeline in social network. 
However, different request classes have diverse latencies. For example, in social network, it takes tens of milliseconds to upload a post, hundreds of milliseconds to update timelines, and a few seconds to perform object detection. To reflect these latency ranges, we assign an SLA per request class and priority. 
We stress test the applications with high user loads and use the latency before saturation as the SLA.
The SLAs of the social network, media service, and video processing pipeline are listed in Table~\ref{tab:sla_social}, \ref{tab:sla_video_share}, \ref{tab:sla_video_pipe}, respectively. 
The SLAs are mostly defined as the $99^{th}$ percentile, except for the low-priority requests in the video processing pipeline, which is defined as the $50^{th}$ percentile latency.

 \begin{figure}[t]
 \centering
 \includegraphics[width=.95\linewidth]{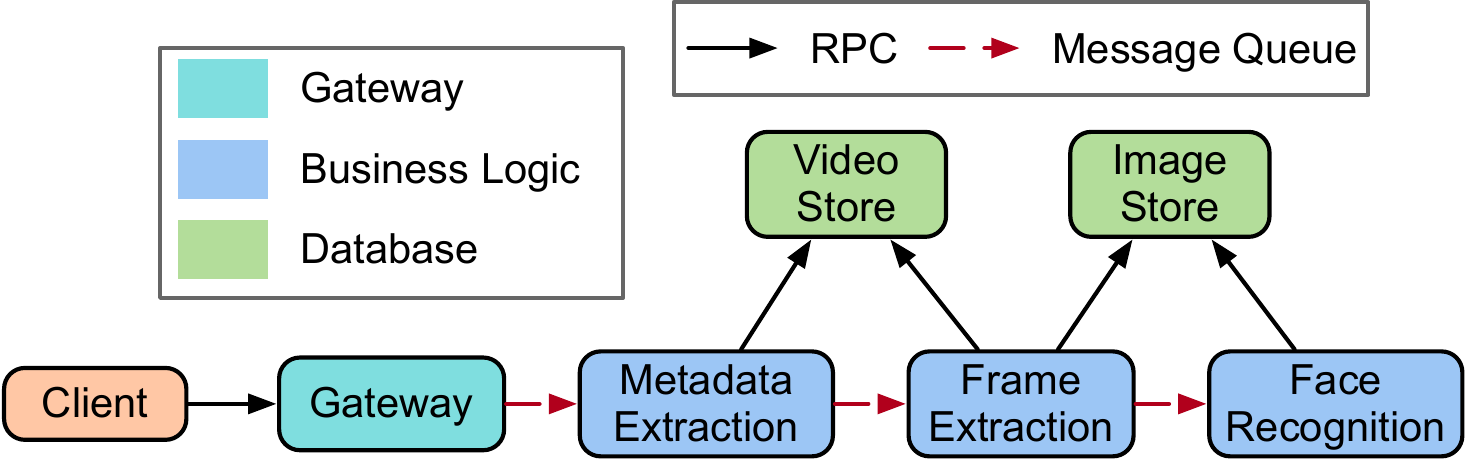}
 \caption{Video processing pipeline.}
 \label{fig:video-pipe}
 \end{figure}

\begin{table}[h]
\centering
\begin{normalsize}
\begin{tabular}{ll}
{\bf Request type}        & {\bf $99^{th}$ latency (ms)} \\ \midrule
upload-post/comment & $75$   \\
read-timeline       & $250$   \\
update-timeline     & $500$   \\
upload-image        & $200$   \\
download-image      & $75$   \\
sentiment-analysis  & $500$   \\
object-detect       & $10000$     \\ \bottomrule  
\end{tabular}
\end{normalsize}
\caption{SLA requirements of the social network.}
\label{tab:sla_social}
\end{table}

\begin{table}[h]
\centering
\begin{normalsize}
\begin{tabular}{ll}
{\bf Request type}        & {\bf $99^{th}$ latency (ms)} \\ \midrule
upload-video        & $2000$   \\
download-video      & $1500$   \\
get-info            & $250$   \\
rate-video          & $400$   \\
transcode-video     & $40000$   \\
generate-thumbnail  & $2000$     \\ \bottomrule  
\end{tabular}
\end{normalsize}
\caption{SLA requirements of the media service.}
\label{tab:sla_video_share}
\end{table}

\begin{table}[h]
\centering
\begin{normalsize}
\begin{tabular}{lll}
{\bf Request type}        & {\bf Percentile} & {\bf Latency (ms)} \\ \midrule
high-priority       & $99^{th}$         & $20000$   \\
low-priority        & $50^{th}$         & $4000$   \\ \bottomrule  
\end{tabular}
\end{normalsize}
\caption{SLA requirements of the video processing pipeline.}
\label{tab:sla_video_pipe}
\end{table}

\section{Evaluation}

We aim to answer the following questions:

\begin{enumerate}
    \item What is the overhead of Ursa's exploration process? (Section~\ref{sec:eval_explore})
    \item How accurate is Ursa's performance model in capturing end-to-end latency (Section~\ref{sec:eval_accuracy})?
    \item How effective is Ursa in reducing resource usage and maintaining SLAs? (Section~\ref{sec:eval_perf})
    \item What is the latency required for Ursa to make resource allocation decisions? (Section~\ref{sec:eval_overhead}) 
    \item Is Ursa able to adapt to business logic changes of microservices? (Section~\ref{sec:eval_online})
\end{enumerate}
\subsection{Experimental Setup}

We use the benchmarks in Section~\ref{sec:benchmark} and use Locust~\cite{locust} to generate input load following a Poisson arrival process. 
The applications are deployed on a local Kubernetes cluster consisting of 8 machines, each with 40-88 CPUs and 126-188 GB of memory each, with a NIC bandwidth of 10 Gbps.
To reduce interference between containers colocated on the same server, we set the CPU management policy of Kubernetes to the static policy~\cite{kbscpu}, which allows each container to access exclusive CPUs, as long as it is configured with an integer number of CPUs. 
The CPU configuration of each microservice's container is determined by monitoring the CPU usage of the container at low RPS and rounding it to the nearest integer, and similarly, the memory configuration is set to the maximum profiled memory usage to avoid OOM errors. During online deployment, we adjust the resource allocation for each microservice by adjusting the number of replicas. 


\subsection{Competing Approaches}

We compare Ursa to the following systems.

\myparagraph{Sinan} Sinan~\cite{zhang2021sinan} is a model-based ML-driven microservice management framework. It uses a CNN and boosted trees model, to predict the end-to-end latency of a microservice topology given a certain resource allocation, as well as the probability that a resource allocation will lead to an SLA violation later into the future, taking into account the system's inertia in building up queues.
Sinan is implemented as a centralized scheduler that periodically queries the model with different resource allocations, and chooses the one using the least amount of resources, while meeting the end-to-end SLA. 
The training data for the models are collected with a process designed to explore unseen resource allocations and to keep the ratio of violating to meeting SLAs at $1:1$, so that the trained models are not biased towards either predicting SLA violation or SLA satisfaction. We train Sinan with 10,000 samples, per the paper's recommendation. 

\myparagraph{Firm}
Firm~\cite{qiu2020firm} is a model-free, ML-driven framework for microservice resource management. Unlike Sinan that trains models to predict latency, Firm assigns a reinforcement learning agent to each service that directly adjusts the resource allocation for the service, given its resource usage and end-to-end SLA status. The reward for each agent is designed to be the weighted sum of the reduced resource usage and the SLA violation status after applying the resource allocation decision.
The agents are trained by injecting performance anomalies during online deployment. Similarly to Sinan, we also use 10,000 training samples for Firm to allow accuracy to converge. 

For both Sinan and Firm, we modified the systems to handle asynchronous events and queues, since the original systems were designed exclusively for RPCs. 


\myparagraph{Autoscaling}
Autoscaling~\cite{aws_step_scaling} is a widely adopted resource management method. The autoscaling controller relies on manually configured resource utilization thresholds based on expert knowledge to dynamically adjust resource allocation. 
In our evaluation, we experiment with two configurations for autoscaling. The first configuration uses the default setting of AWS step scaling~\cite{aws_step_scaling}, which increases resources when CPU utilization exceeds 60\%, and reduces resources when the CPU utilization is below 30\%. This configuration is optimistic in terms of resource usage, but comes at the expense of SLA maintenance.
The second configuration is manually tuned to preserve the SLAs of tested applications, 
but uses more resources. In the rest of the paper, we name the first configuration Auto-a, and the second configuration Auto-b.
\subsection{Exploration Overhead}
\label{sec:eval_explore}

We now compare the exploration overheads of Ursa, Sinan and Firm. During exploration, the combination of user requests for each application is the same across the three approaches. 
Specifically, for the social network application, the ratios of post, comment, download-image and read-timeline are approximately 1:75:15:25, adopted from~\cite{zhang2021sinan, kwak2010twitter, gligoric2018constraints}. For the media service application, the ratios of upload-video, get-info, download-video, and rate-video are approximately 1:100:25:25. For the video processing pipeline, we experiment with four different ratios of high and low priority requests, including 5:95, 25:75, 50:50, and 75:25. 
Across all approaches, the sampling frequency is set to once per minute.

We run Ursa's exploration process for each microservice individually, as described in Algorithm~\ref{algo:service_exploration}.
Since CPU and memory are the two major resources on cloud platforms, where CPU directly affects latency and memory is configured to avoid OOM error, during Ursa's exploration we configure the initial replica of each microservice by providing it with adequate CPUs to keep the microservice's latency low. 
Each microservice is profiled using Algorithm~\ref{algo:service_exploration}, and in each iteration we reduce the number of replicas by 1 and collect 10 samples, until the frequency of SLA violations exceeds $10\%$, or the CPU utilization exceeds the backpressure-free threshold.
For Sinan and Firm, we run their data collection algorithm and online training process separately and collect 10k samples for each application, matching the order of magnitude in Sinan for DeathStarBench~\cite{zhang2021sinan}. 

Table~\ref{tab:explore_overhead} summarizes the number of samples collected during exploration and the required exploration time. For Ursa, the exploration time is decided by the longest time required to profile a single microservice, as each microservice can be profiled individually, and the number of samples is the sum of samples collected for all microservices. 
Compared to the ML-driven approaches, Ursa reduces the required sample size by a factor of $16.7$ up to $25.6$, and the exploration time by a factor of $128.2$ up to $208.4$. 
The high online exploration overheads of ML-driven approaches result from the nature of deep neural networks, which require a large amount of data to generalize, due to their large parameter space. 
In contrast, Ursa's analytical model inherently contains fewer parameters than DNNs. The model calculates end-to-end latency as the sum of per-microservice latencies, and foresees end-to-end SLA violations when the latency of individual microservices increases rapidly. 
Notably, despite the small required sample size, Ursa still maintains SLA and achieves high efficiency during deployment, as shown in Section~\ref{sec:eval_perf}.


\begin{table}[h]
\centering
\begin{normalsize}
\begin{tabular}{llll}
App                       & Systems    & Samples   & Time(h)   \\ \midrule
\multirow{2}{*}{Social}   & Ursa      & 440       & 1.2              \\
                                  & Sinan/Firm      & 10000     & 166.7     \\ \midrule 
\multirow{2}{*}{Media}    & Ursa      & 390       & 0.8              \\
                                  & Sinan/Firm      & 10000     & 166.7    \\ \midrule 
\multirow{2}{*}{Video} & Ursa      & 600       & 0.8      \\
                                  & Sinan/Firm      & 10000     & 166.7   \\ \bottomrule 
\end{tabular}
\end{normalsize}
\caption{Exploration overheads with Ursa compared to two ML-driven frameworks, Sinan and Firm. 
}
\label{tab:explore_overhead}
\end{table}
\subsection{Model Accuracy}
\label{sec:eval_accuracy}

The performance model (Section~\ref{sec:performance_model}) estimates the end-to-end latency by multiplying the latency's upper bound with the expected overestimation rate. 
To evaluate the estimation accuracy, we record the per-microservice and end-to-end latency distributions every 5 min for a total of 150 min during online exploration with dynamically changing resource allocations, and calculate the estimated latency for each type of request.

\begin{figure}[t]
\centering
\includegraphics[width=1.0\linewidth]{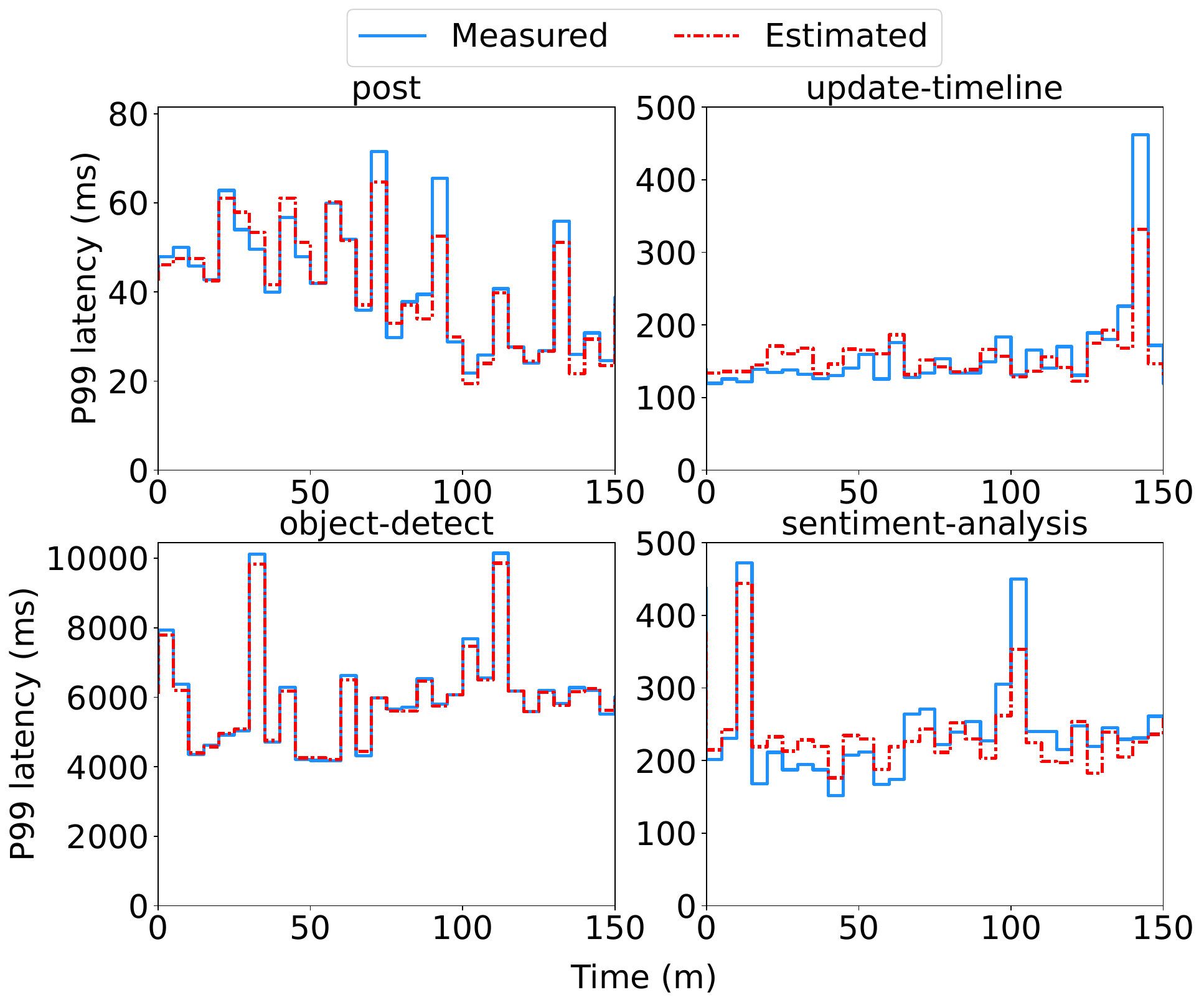}
 \vspace{-0.26in}
\caption{Social network--Estimated vs. measured latency.
}
\label{fig:acc_social}
\end{figure}

\begin{figure}[t]
\centering
\includegraphics[width=.97\linewidth]{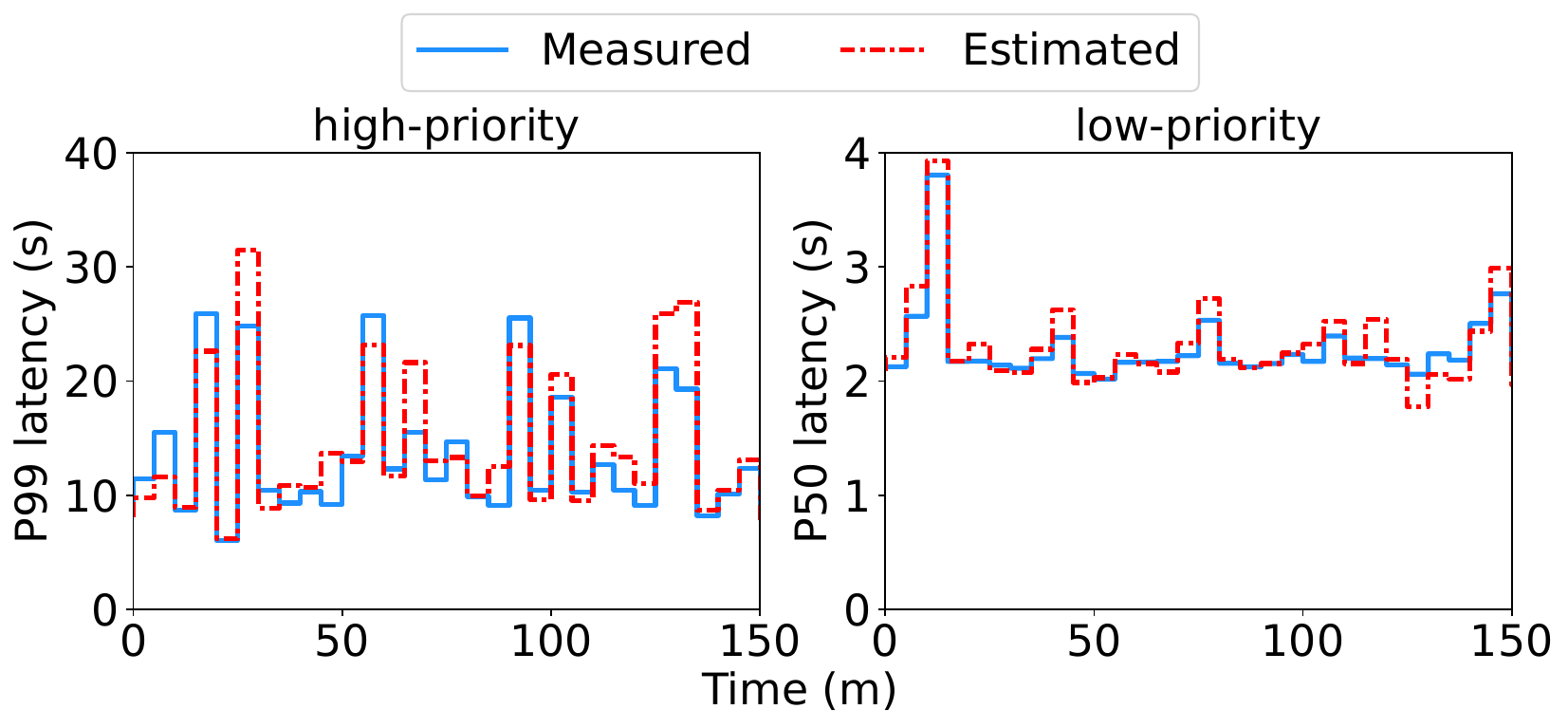}
\caption{Estimated vs. measured latency for the video processing pipeline.
}
\label{fig:acc_vpipe}
\end{figure}

Figure~\ref{fig:acc_social} shows the measured and estimated latency of four representative request types in Social Network, including post, update-timeline, object-detection, and sentiment-analysis. The blue line indicates the measured $99^{th}$ percentile latency and the red line indicates the estimated $99^{th}$ percentile latency. 
For each class of requests, the estimated latency closely follows the measured latency, with the average ratio of estimated to measured latency ranging from $0.97$ to $1.05$.
Additionally, Figure~\ref{fig:acc_vpipe} shows the measured and estimated latency of the video processing pipeline which includes two request priorities, with SLAs defined at the $50^{th}$ and $99^{th}$ percentiles, for low and high priority requests, respectively. For both priorities, the estimated latency is close to the measured latency, with the average ratio of estimated to measured latency being $0.96$ and $1.00$ for low and high priority requests, respectively.

\subsection{Performance Comparison}
\label{sec:eval_perf}

\begin{figure*}[t]
  \centering
    \includegraphics[width=0.8\linewidth]{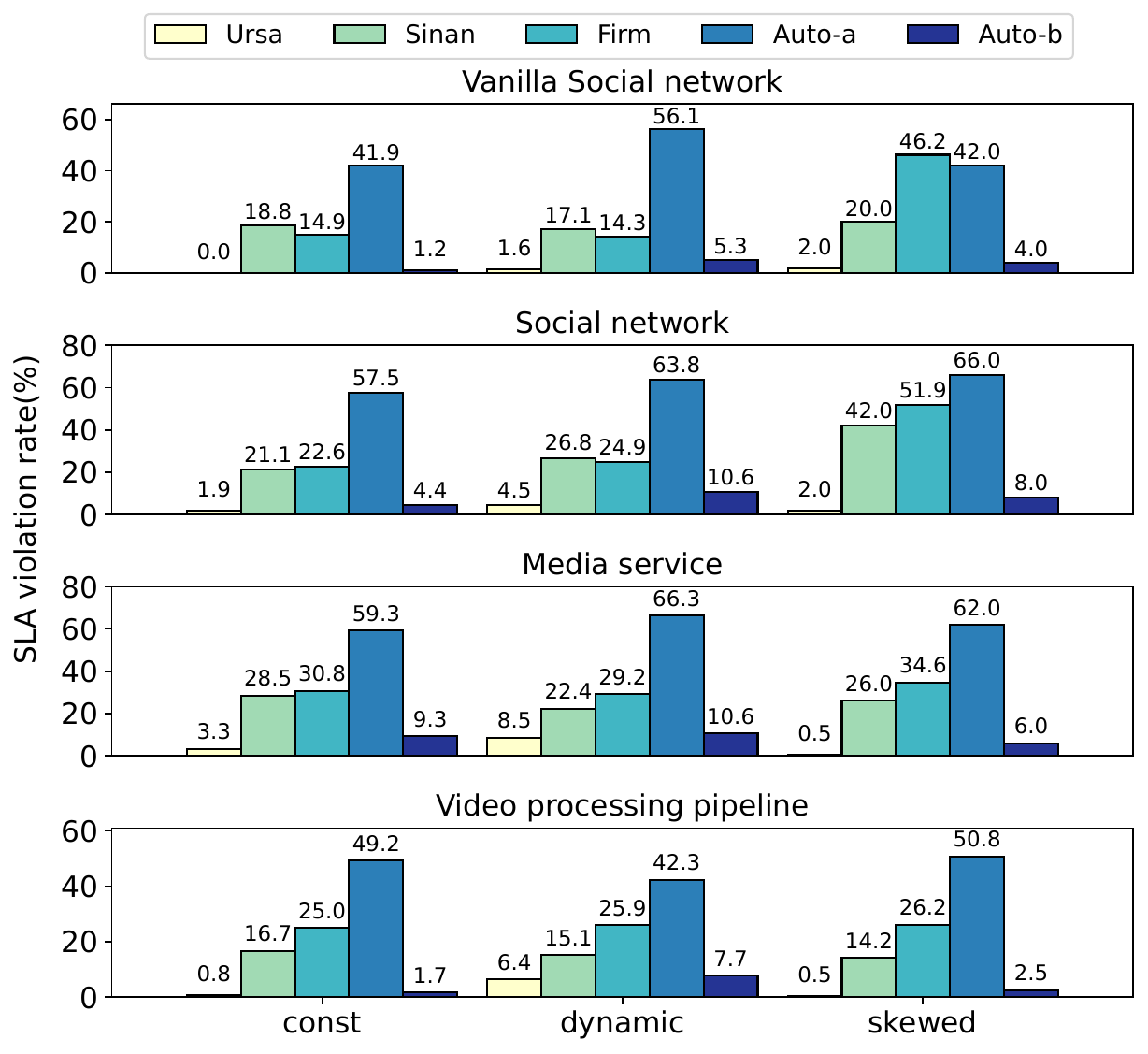}
\caption{SLA violation rate across load patterns for the original (Vanilla) and reimplemented Social Network application, the Media service, and the Video Processing pipeline, with Ursa, Sinan, Firm, and the empirical Automscaling system.}
\label{fig:sla_viol}
\end{figure*} 

We now compare resource usage and SLA violations during deployment between Ursa and prior work, with Ursa and ML-driven systems using exploration data from Section~\ref{sec:eval_explore}.
In addition to the three previous applications, we also show the results for the vanilla social network which implements the same functionality as the original benchmark, by disabling the newly added ML services, to highlight the challenges that stem from resource need heterogeneity across microservices.

For each application, we experiment with three user loads; \textit{constant load}, \textit{dynamic load}, and \textit{skewed load}. 
\textit{Constant load} refers to Poisson arrival processes with constant RPS.
\textit{Dynamic load} has time-varying RPS, including diurnal patterns where the RPS first gradually increases and then gradually decreases, and burst patterns, where the RPS increases sharply by 50\% to 125\%. The ratio of different types of requests for constant and dynamic loads is the same as in online exploration. 
In \textit{skewed load}, the ratio of request types differs from that in the online exploration. 
For social network and media service, we experiment with two other request combinations, the first doubling the frequency of update requests, 
and the second halving the frequency of update requests. 
For the video processing pipeline, the ratios of high-priority to low-priority requests include 40:60 and 60:40, which do not exist in online exploration. 
For Ursa specifically, the skewed load stresses the case where the request mix changes, and the LPR thresholds needs recalculated using available exploration data that do not include the current request mix. For each type of load, Ursa calculates the optimal load-per-replica thresholds once, at the beginning of the experiment.

\begin{figure*}[t]
  \centering
    \includegraphics[width=0.8\linewidth]{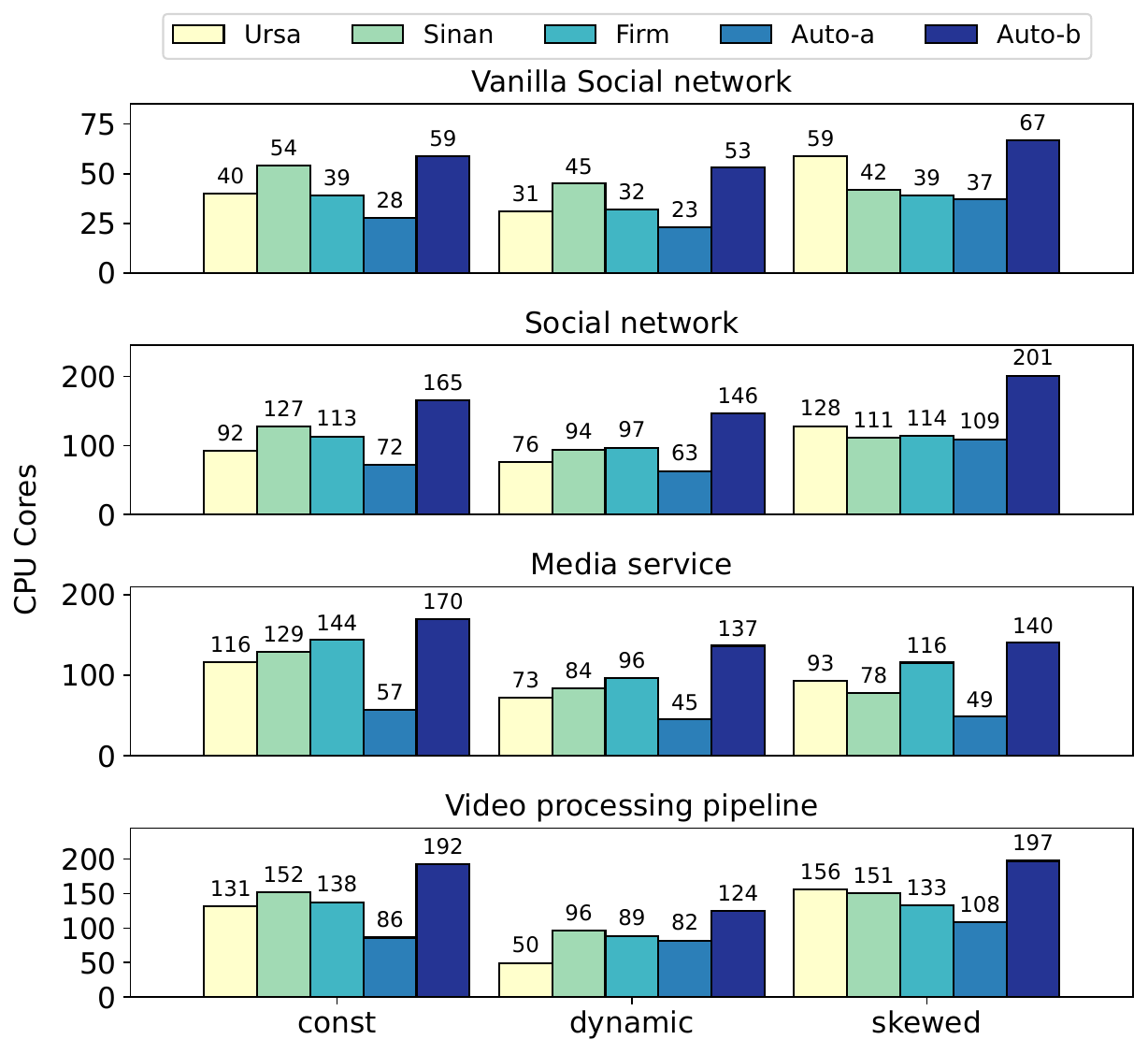}
\caption{Average CPU allocation across load patterns for the original (Vanilla) and reimplemented Social Network application, the Media service, and the Video Processing pipeline, with Ursa, Sinan, Firm, and the empirical Automscaling system. 
}
\label{fig:cpu_usage}
\end{figure*}

Figure~\ref{fig:sla_viol} shows the SLA violation rate, and Figure~\ref{fig:cpu_usage} shows the average CPU allocation, \camr{or the total amount of CPU resources allocated to the microservice}. 
Compared to ML-driven systems, Ursa significantly reduces SLA violation rates, achieving $0.1\%$ to $8.5\%$ SLA violation rates under constant and dynamic loads, and $0.5\%$ to $2.0\%$ SLA violation rates under skewed load, whereas ML-driven systems incur $9.1\%$ to $29.2\%$ SLA violation rates under constant and dynamic loads, and $14.2\%$ to $51.9\%$ SLA violation rates under skewed load.
ML-driven systems cause higher SLA violation rates for the new social network than for vanilla social network, because the latency of ML microservices is less stable and more challenging for resource management, compared to lightweight text processing.
In terms of resources, Ursa reduces CPU allocation by $2.3\%$ to $86.2\%$ for constant and dynamic loads. For skewed loads, Ursa uses an average of $8.2\%$ more CPUs, but the ML-driven systems result in SLA violation rates significantly higher than Ursa. 
As for autoscaling, Auto-a, which uses the default setting of AWS step scaling, uses the least resources but results in SLA violation rates of over $40\%$. Auto-b, the manually tuned configuration, maintains SLAs most of the time, with SLA violation rates only $0.9\%$ to $6.1\%$ higher than Ursa. Despite the low SLA violation rates, Auto-b uses significantly more resources than Ursa, allocating $43.9\%$ to $148.0\%$ more CPUs under constant and dynamic loads, and $13.6\%$ to $57.0\%$ more CPUs allocation under skewed loads.

\begin{figure*}[t]
  \centering
    \includegraphics[width=0.74\linewidth]{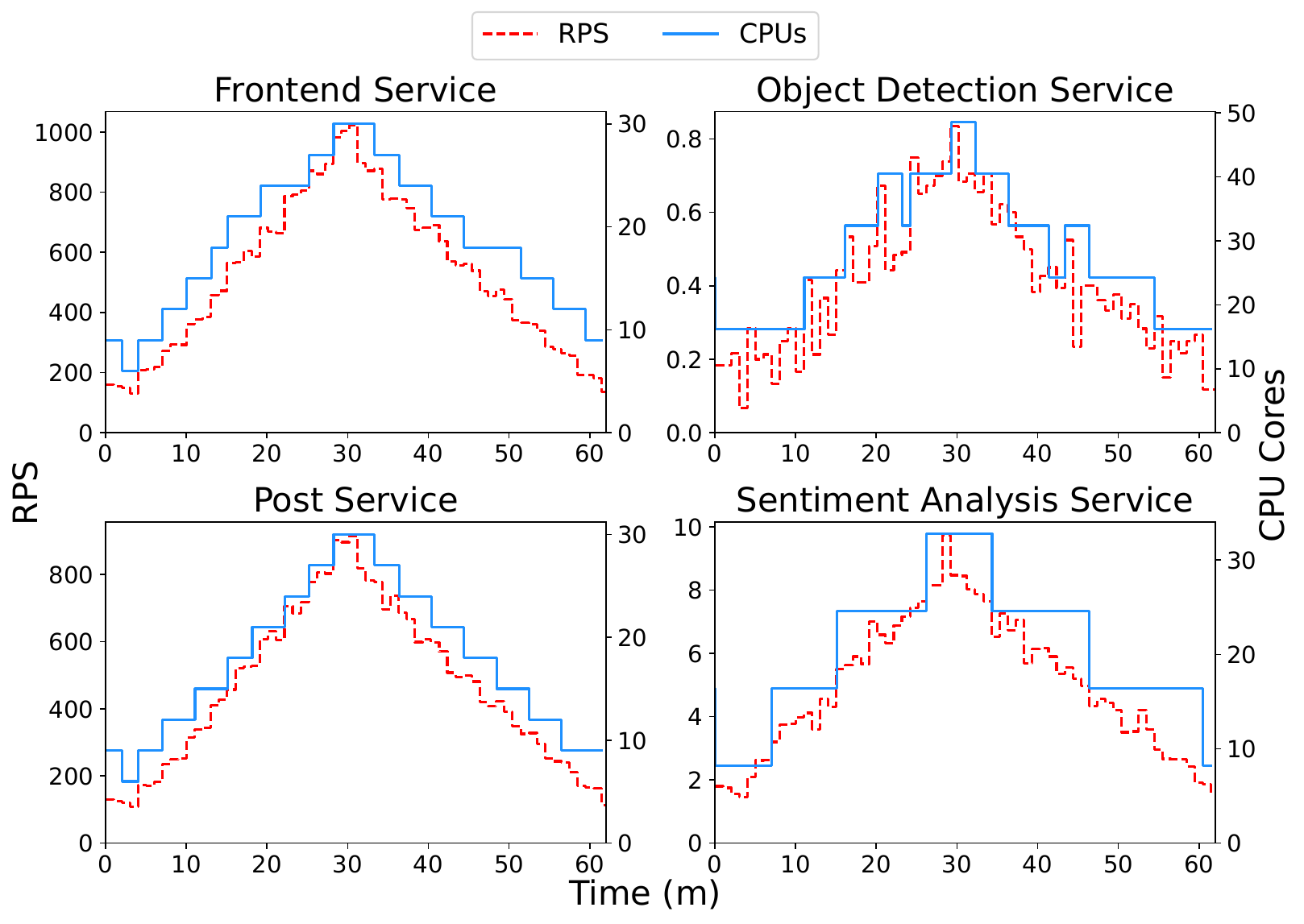}
\caption{Ursa's CPU allocation under a diurnal load. We show the RPS and CPU allocations for individual, representative microservices. 
}
\label{fig:social_diurnal}
\end{figure*}

Ursa may use more resources under skewed loads because it prioritizes maintaining SLAs and makes conservative decisions with the available exploration data. 
As a conceptual example, assume a microservice handles two classes of requests, and its total load is $(4, 6)$, where each element of the vector is the load of one class of requests. If the microservice's exploration data only includes one feasible LPR threshold $(3, 2)$, Ursa will provision 3 replicas for the microservice to ensure that the load of any request class is below the threshold, while in reality the actual per-replica load will be $(1.3, 2)$, which is below the $(3, 2)$ threshold.
Figure~\ref{fig:social_diurnal} shows the load and CPU allocation for four representative microservices in the social network under a diurnal load when managed with Ursa, where the left Y-axis represents the RPS of load and the right Y-axis represents the CPU allocation. For each microservice, Ursa is able to scale out and scale in promptly as the load increases and decreases.

Ursa outperforms ML-driven systems with much lower exploration overheads, because Ursa's analytical model accurately decomposes the end-to-end latency to per-microservice latencies, which can be mapped directly to per-microservice resource allocation.
However, the ML-driven techniques need to learn the relation between resource allocation and SLA from scratch in a much larger parameter space, requiring more data and leading to lower accuracy. Specifically, Sinan's SLA violation predictor can only achieve $80\%$ to $85\%$ accuracy due to the presence of multiple request classes with different SLAs, resulting in more SLA violations and higher resource usage. 
On the other hand, in addition to the issue of large parameter space, Firm does not always prioritize preserving SLAs because its agent's reward function is a weighted sum of the SLA violation rate and the resource utilization, which makes Firm prioritize resource savings over SLA if the savings are significant, resulting in more violations. 

\camr{
\textbf{Discussion.} We previously compared the resource consumption of the evaluated systems. Another metric of interest in cloud native environments is throughput per dollar, i.e., the user request throughput achievable with the same cost budget. Since all systems are evaluated under the same workload patterns, the improvement in Ursa's throughput per dollar is the inverse of its resource savings. For example, compared to ML-driven systems, Ursa reduces CPU allocation by $2.3\%$ to $86.2\%$ for constant and dynamic loads, which represents an improvement in throughput per dollar of $1.02\times$ to $7.24\times$. Ursa's improvement is even more pronounced when considering goodput per dollar, i.e., the user request throughput that meets SLA under the same cost budget, since Ursa significantly reduces SLA violations compared to other systems.
}

\subsection{Control Plane Latency}
\label{sec:eval_overhead}

The latency of resource allocation decisions is directly related to responsiveness of resource management systems.
Resource allocation decisions are fast in Ursa because the critical path only includes the resource controller, which calculates the number of replicas, based on the load-per-replica thresholds, while inference of ML requires over thousands of floating-point operations. 
There are also situations where the model needs to be updated to account for changes in service business logic or load combinations. In such cases, Ursa needs to recompute the optimization models, and ML-driven approaches also need to be retrained.
Table~\ref{tab:ctrl_plane_latency} shows the average control plane latency (in milliseconds) across the different approaches, in the case of deployment and model update. In the comparison, the control planes are always allocated 4 CPUs. 
Autoscaling is undoubtedly the fastest, as it involves only a single threshold check. In terms of deployment, Ursa is on average $691.6\times$ faster than Sinan with its centralized ML model, and $43.4\times$ faster than Firm, which uses per-service RL agents. 
In terms of model updates, Sinan's retraining time is linear with the size of the dataset, and takes minutes even on a dedicated GPU. Firm can adapt to load changes gradually by updating the RL agent online, but is still slower than Ursa by $4.4\times$ even for a single iteration.
The RL agent may require thousands of iterations to update its weights and fully learn new resource patterns, whereas Ursa only needs to solve the optimization problem once to fully adapt to the changes. 

\begin{table}[ht]
\centering
\begin{normalsize}
\begin{tabular}{lllll}
                    & {\bf Ursa}  & {\bf Sinan} & {\bf Firm} & {\bf Autoscaling} \\ \midrule
{\bf Deploy}              & 0.5   & 345.8 & 21.7 & 0.1     \\
{\bf Update}              & 271.7 & N/A   & 1.2 $\times 10^3$ & 0.1     \\ \bottomrule  
\end{tabular}
\end{normalsize}
\caption{Average control plane latency in milliseconds across Ursa, Sinan, Firm, and Autoscaling for the initial application deployment, as well as when retraining of the model is required, due to a change in application logic. 
}
\label{tab:ctrl_plane_latency}
\end{table}

\subsection{Adapting to Service Changes}
\label{sec:eval_online}
\begin{figure}[t]
  \centering
    \includegraphics[width=.7\linewidth]{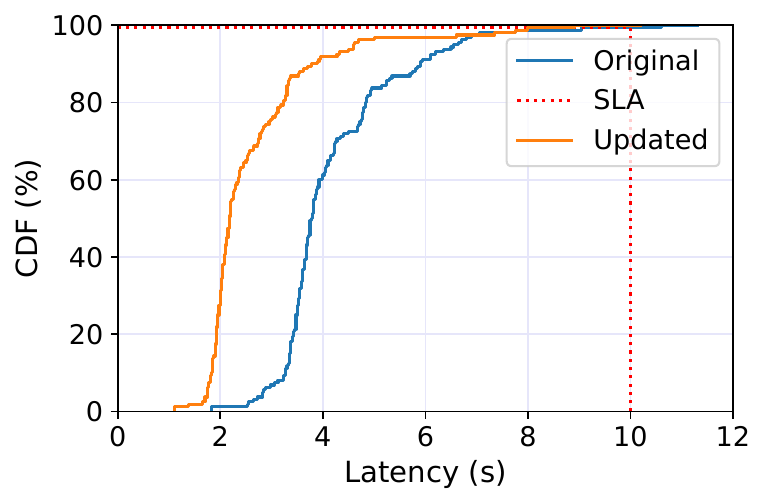}
\caption{$99^{th}$ latency distribution of object-detect.}
\label{fig:online_object_detect}
\end{figure}

\camr{
A basic premise of microservices is that they enable frequent logic updates without the high overhead of redeploying the entire service mesh. We now conduct a case study to demonstrate Ursa's ability to adapt to such business logic updates. Specifically, we modify the object-detection service in the social network application, and change the model 
from DETR~\cite{carion2020end} which combines Transformer~\cite{vaswani2017attention} and Resnet~\cite{he2016deep} for object detection, 
to the more lightweight Mobilenet~\cite{howard2017mobilenets}. 
The exploration controller performs a partial online exploration to profile only the modified object-detect service. It collects a total of 75 samples in 1.25 hours, during which 4 SLA violations are triggered, resulting in an SLA violation rate of $5.3\%$. 
Then the optimization engine recalculates the LPR threshold of each service. 
We deploy the modified social network application under various RPS, and Figure~\ref{fig:online_object_detect} shows the distribution of the $99^{th}$ percentile latency of the end-to-end object-detect requests for the original and the updated Social network service mesh. 
Object-detect requests go through the frontend service, image store, post service, and the object-detect service. 
The red line represents the SLA and the blue line represents the cumulative distribution function, with SLA violation rates of $0.62\%$ and $0.50\%$ for the original and updated microservice, respectively, showing Ursa's ability to quickly adjust to changes in the application logic. 
}

\subsection{Summary}

Compared to ML-driven approaches, Ursa reduces the required exploration time by more than $128 \times$, making it practical to track frequent changes in service business logic and user loads. Ursa also achieves better performance, maintaining low SLA violation rates of $0.1\%$ to $8.5\%$ during deployment, $9.0\%$ to $49.9\%$ lower than ML-driven approaches, and reducing resource allocation by up to $86.2\%$. Ursa also outperforms traditional autoscaling, both the default setting of AWS step scaling, and a conservative, manually tuned configuration. 
In addition, Ursa's control plane is faster than ML, enabling faster and more scalable management decisions. 
Finally, we demonstrate that Ursa is able to adapt to service changes and maintain SLAs while incurring low exploration overheads.

\section{Conclusion}

We present Ursa, a lightweight resource management framework for microservices. Ursa uses an analytical model to decompose the end-to-end SLA into per-microservice SLAs, and maps them to resource allocations. During exploration, Ursa profiles each microservice individually and swiftly stops exploration in the case of SLA violations to shorten the exploration process. Using benchmarks implemented with popular microservice frameworks, we demonstrate that Ursa outperforms ML-driven approaches and traditional autoscaling in both SLA maintenance and resource efficiency, with significantly lower exploration overheads.

\section*{Acknowledgements}

We sincerely thank Ricardo Bianchini, Rodrigo Fonseca, \'I\~{n}igo Goiri, Mahesh Ketkar, Ramesh Illikkal, Nikita Lazarev, Mingyu Liang, Varun Gohil, and the anonymous reviewers for their feedback on earlier versions of this manuscript. This work was partly done during an internship at Microsoft Research and was in part supported by an NSF CAREER Award CCF-1846046, an  Intel Research Award, an Intel Faculty Rising Star Award, a Sloan Research Fellowship, a Microsoft Research Fellowship, a Facebook Research Faculty Award, and a Google Research Award. 


\bibliographystyle{IEEEtranS}
\bibliography{references}

\begin{thebibliography}{10}
\providecommand{\url}[1]{#1}
\csname url@samestyle\endcsname
\providecommand{\newblock}{\relax}
\providecommand{\bibinfo}[2]{#2}
\providecommand{\BIBentrySTDinterwordspacing}{\spaceskip=0pt\relax}
\providecommand{\BIBentryALTinterwordstretchfactor}{4}
\providecommand{\BIBentryALTinterwordspacing}{\spaceskip=\fontdimen2\font plus
\BIBentryALTinterwordstretchfactor\fontdimen3\font minus \fontdimen4\font\relax}
\providecommand{\BIBforeignlanguage}[2]{{%
\expandafter\ifx\csname l@#1\endcsname\relax
\typeout{** WARNING: IEEEtranS.bst: No hyphenation pattern has been}%
\typeout{** loaded for the language `#1'. Using the pattern for}%
\typeout{** the default language instead.}%
\else
\language=\csname l@#1\endcsname
\fi
#2}}
\providecommand{\BIBdecl}{\relax}
\BIBdecl

\bibitem{ack}
``{Alibaba Cloud Container Service for Kubernetes},'' \url{https://www.alibabacloud.com/product/kubernetes}.

\bibitem{eks}
``{Amazon Elastic Kubernetes Service},'' \url{https://aws.amazon.com/eks/}.

\bibitem{kafka}
``{Apache Kafka},'' \url{https://kafka.apache.org/}.

\bibitem{aks}
``{Azure Kubernetes Service},'' \url{https://azure.microsoft.com/en-us/products/kubernetes-service/#overview}.

\bibitem{dapr}
``{Dapr: APIs for building portable and reliable microservices},'' \url{https://dapr.io/}.

\bibitem{twitter_decomposing}
``Decomposing twitter: Adventures in service-oriented architecture,'' {\url{https://www.slideshare.net/InfoQ/decomposing-twitter-adventures-in-serviceoriented-architecture}}.

\bibitem{ffmpeg}
``{FFmpeg: A complete, cross-platform solution to record, convert and stream audio and video},'' \url{https://ffmpeg.org/}.

\bibitem{gke}
``{Google Kubernetes Engine},'' \url{https://cloud.google.com/kubernetes-engine}.

\bibitem{gRPC}
``{gRPC: A high performance, open source universal RPC framework},'' \url{https://grpc.io/}.

\bibitem{gurobi}
``{Gurobi Optimization},'' \url{https://www.gurobi.com/}.

\bibitem{huggingface}
``{Hugging Face},'' \url{https://huggingface.co/}.

\bibitem{k8s}
``{Kubernetes: Production-Grade Container Orchestration},'' \url{https://kubernetes.io/}.

\bibitem{locust}
``Locust: A modern load testing framework,'' \url{https://locust.io/}.

\bibitem{opencv}
``{OpenCV Face Recognition},'' \url{https://opencv.org/}.

\bibitem{prometheus}
``{Prometheus},'' \url{https://prometheus.io/}.

\bibitem{redis}
``{Redis},'' \url{https://redis.io/}.

\bibitem{redis-streams}
``{Redis streams tutorial},'' \url{https://redis.io/docs/data-types/streams-tutorial/}.

\bibitem{aws_step_scaling}
``Step and simple scaling policies for amazon ec2 auto scaling,'' \url{https://docs.aws.amazon.com/autoscaling/ec2/userguide/as-scaling-simple-step.html}.

\bibitem{ambati2020providing}
P.~Ambati, {\'I}.~Goiri, F.~Frujeri, A.~Gun, K.~Wang, B.~Dolan, B.~Corell, S.~Pasupuleti, T.~Moscibroda, S.~Elnikety, and R.~Bianchini, ``{Providing SLOs for Resource-Harvesting VMs in Cloud Platforms},'' in \emph{OSDI}, 2020.

\bibitem{boutin2014apollo}
E.~Boutin, J.~Ekanayake, W.~Lin, B.~Shi, J.~Zhou, Z.~Qian, M.~Wu, and L.~Zhou, ``{Apollo: Scalable and coordinated scheduling for cloud-scale computing},'' in \emph{OSDI}, 2014.

\bibitem{carion2020end}
N.~Carion, F.~Massa, G.~Synnaeve, N.~Usunier, A.~Kirillov, and S.~Zagoruyko, ``End-to-end object detection with transformers,'' in \emph{European conference on computer vision}.\hskip 1em plus 0.5em minus 0.4em\relax Springer, 2020, pp. 213--229.

\bibitem{chow2022deeprest}
K.-H. Chow, U.~Deshpande, S.~Seshadri, and L.~Liu, ``Deeprest: deep resource estimation for interactive microservices,'' in \emph{Proceedings of the Seventeenth European Conference on Computer Systems}, 2022, pp. 181--198.

\bibitem{cortez2017resource}
E.~Cortez, A.~Bonde, A.~Muzio, M.~Russinovich, M.~Fontoura, and R.~Bianchini, ``Resource central: Understanding and predicting workloads for improved resource management in large cloud platforms,'' in \emph{Proceedings of the 26th Symposium on Operating Systems Principles}.\hskip 1em plus 0.5em minus 0.4em\relax ACM, 2017, pp. 153--167.

\bibitem{delgado2015hawk}
P.~Delgado, F.~Dinu, A.-M. Kermarrec, and W.~Zwaenepoel, ``{Hawk: Hybrid datacenter scheduling},'' in \emph{USENIX ATC}, 2015.

\bibitem{Delimitrou13}
C.~Delimitrou and C.~Kozyrakis, ``{Paragon: QoS-Aware Scheduling for Heterogeneous Datacenters},'' in \emph{Proceedings of the Eighteenth International Conference on Architectural Support for Programming Languages and Operating Systems (ASPLOS)}.\hskip 1em plus 0.5em minus 0.4em\relax Houston, TX, USA, 2013.

\bibitem{Delimitrou13d}
C.~Delimitrou and C.~Kozyrakis, ``{QoS-Aware Scheduling in Heterogeneous Datacenters with Paragon},'' in \emph{ACM Transactions on Computer Systems (TOCS), Vol. 31 Issue 4}.\hskip 1em plus 0.5em minus 0.4em\relax December 2013.

\bibitem{Delimitrou14b}
C.~Delimitrou and C.~Kozyrakis, ``{Quality-of-Service-Aware Scheduling in Heterogeneous Datacenters with Paragon},'' in \emph{IEEE Micro Special Issue on Top Picks from the Computer Architecture Conferences}.\hskip 1em plus 0.5em minus 0.4em\relax May/June 2014.

\bibitem{Delimitrou14}
C.~Delimitrou and C.~Kozyrakis, ``{Quasar: Resource-Efficient and QoS-Aware Cluster Management},'' in \emph{Proceedings of the Nineteenth International Conference on Architectural Support for Programming Languages and Operating Systems (ASPLOS)}.\hskip 1em plus 0.5em minus 0.4em\relax Salt Lake City, UT, USA, 2014.

\bibitem{Delimitrou16}
C.~Delimitrou and C.~Kozyrakis, ``{HCloud: Resource-Efficient Provisioning in Shared Cloud Systems},'' in \emph{Proceedings of the Twenty First International Conference on Architectural Support for Programming Languages and Operating Systems (ASPLOS)}, April 2016.

\bibitem{Delimitrou17}
C.~Delimitrou and C.~Kozyrakis, ``{Bolt: I Know What You Did Last Summer... In The Cloud},'' in \emph{Proc. of the Twenty Second International Conference on Architectural Support for Programming Languages and Operating Systems (ASPLOS)}, 2017.

\bibitem{Delimitrou15}
C.~Delimitrou, D.~Sanchez, and C.~Kozyrakis, ``{Tarcil: Reconciling Scheduling Speed and Quality in Large Shared Clusters},'' in \emph{Proceedings of the Sixth ACM Symposium on Cloud Computing (SOCC)}, August 2015.

\bibitem{ferguson2012jockey}
A.~D. Ferguson, P.~Bodik, S.~Kandula, E.~Boutin, and R.~Fonseca, ``{Jockey: Guaranteed Job Latency in Data Parallel Clusters},'' in \emph{EuroSys}, 2012.

\bibitem{Gan18}
Y.~Gan and C.~Delimitrou, ``{The Architectural Implications of Cloud Microservices},'' in \emph{Computer Architecture Letters (CAL), vol.17, iss. 2}, Jul-Dec 2018.

\bibitem{gan2021sage}
Y.~Gan, M.~Liang, S.~Dev, D.~Lo, and C.~Delimitrou, ``Sage: practical and scalable ml-driven performance debugging in microservices,'' in \emph{Proceedings of the 26th ACM International Conference on Architectural Support for Programming Languages and Operating Systems}, 2021, pp. 135--151.

\bibitem{Gan18b}
Y.~Gan, M.~Pancholi, D.~Cheng, S.~Hu, Y.~He, and C.~Delimitrou, ``{Seer: Leveraging Big Data to Navigate the Complexity of Cloud Debugging},'' in \emph{Proceedings of the Tenth USENIX Workshop on Hot Topics in Cloud Computing (HotCloud)}, July 2018.

\bibitem{gan2019open}
Y.~Gan, Y.~Zhang, D.~Cheng, A.~Shetty, P.~Rathi, N.~Katarki, A.~Bruno, J.~Hu, B.~Ritchken, B.~Jackson, K.~Hu, M.~Pancholi, B.~Clancy, C.~Colen, F.~Wen, C.~Leung, S.~Wang, L.~Zaruvinsky, M.~Espinosa, Y.~He, and C.~Delimitrou, ``An open-source benchmark suite for microservices and their hardware-software implications for cloud \& edge systems,'' in \emph{Proceedings of the Twenty-Fourth International Conference on Architectural Support for Programming Languages and Operating Systems}.\hskip 1em plus 0.5em minus 0.4em\relax ACM, 2019, pp. 3--18.

\bibitem{ghodsi2011dominant}
A.~Ghodsi, M.~Zaharia, B.~Hindman, A.~Konwinski, S.~Shenker, and I.~Stoica, ``{Dominant Resource Fairness: Fair Allocation of Multiple Resource Types},'' in \emph{NSDI}, 2011.

\bibitem{gligoric2018constraints}
K.~Gligori{\'c}, A.~Anderson, and R.~West, ``How constraints affect content: The case of twitter’s switch from 140 to 280 characters,'' in \emph{Twelfth International AAAI Conference on Web and Social Media}, 2018.

\bibitem{gog2016firmament}
I.~Gog, M.~Schwarzkopf, A.~Gleave, R.~N. Watson, and S.~Hand, ``{Firmament: Fast, centralized cluster scheduling at scale},'' in \emph{OSDI}, 2016.

\bibitem{he2016deep}
K.~He, X.~Zhang, S.~Ren, and J.~Sun, ``Deep residual learning for image recognition,'' in \emph{Proceedings of the IEEE conference on computer vision and pattern recognition}, 2016, pp. 770--778.

\bibitem{howard2017mobilenets}
A.~G. Howard, M.~Zhu, B.~Chen, D.~Kalenichenko, W.~Wang, T.~Weyand, M.~Andreetto, and H.~Adam, ``Mobilenets: Efficient convolutional neural networks for mobile vision applications,'' \emph{arXiv preprint arXiv:1704.04861}, 2017.

\bibitem{isard2009quincy}
M.~Isard, V.~Prabhakaran, J.~Currey, U.~Wieder, K.~Talwar, and A.~Goldberg, ``{Quincy: fair scheduling for distributed computing clusters},'' in \emph{SOSP}, 2009.

\bibitem{jyothi2016morpheus}
S.~A. Jyothi, C.~Curino, I.~Menache, S.~M. Narayanamurthy, A.~Tumanov, J.~Yaniv, R.~Mavlyutov, I.~Goiri, S.~Krishnan, J.~Kulkarni \emph{et~al.}, ``{Morpheus: Towards automated slos for enterprise clusters},'' in \emph{SOSP}, 2016.

\bibitem{karanasos2015mercury}
K.~Karanasos, S.~Rao, C.~Curino, C.~Douglas, K.~Chaliparambil, G.~M. Fumarola, S.~Heddaya, R.~Ramakrishnan, and S.~Sakalanaga, ``{Mercury: Hybrid centralized and distributed scheduling in large shared clusters},'' in \emph{USENIX ATC}, 2015.

\bibitem{kbscpu}
{Kubernetes}, ``Kubernetes cpu management policy,'' \url{https://kubernetes.io/docs/tasks/administer-cluster/cpu-management-policies/}, 2021.

\bibitem{kwak2010twitter}
H.~Kwak, C.~Lee, H.~Park, and S.~Moon, ``What is twitter, a social network or a news media?'' in \emph{Proceedings of the 19th international conference on World wide web}.\hskip 1em plus 0.5em minus 0.4em\relax AcM, 2010, pp. 591--600.

\bibitem{land2010automatic}
A.~H. Land and A.~G. Doig, ``An automatic method for solving discrete programming problems,'' in \emph{50 Years of Integer Programming 1958-2008}.\hskip 1em plus 0.5em minus 0.4em\relax Springer, 2010, pp. 105--132.

\bibitem{luo2021characterizing}
S.~Luo, H.~Xu, C.~Lu, K.~Ye, G.~Xu, L.~Zhang, Y.~Ding, J.~He, and C.~Xu, ``Characterizing microservice dependency and performance: Alibaba trace analysis,'' in \emph{Proceedings of the ACM Symposium on Cloud Computing}, 2021, pp. 412--426.

\bibitem{luo2022power}
S.~Luo, H.~Xu, K.~Ye, G.~Xu, L.~Zhang, G.~Yang, and C.~Xu, ``The power of prediction: microservice auto scaling via workload learning,'' in \emph{Proceedings of the 13th Symposium on Cloud Computing}, 2022, pp. 355--369.

\bibitem{mahgoub2022orion}
A.~Mahgoub, E.~B. Yi, K.~Shankar, S.~Elnikety, S.~Chaterji, and S.~Bagchi, ``$\{$ORION$\}$ and the three rights: Sizing, bundling, and prewarming for serverless $\{$DAGs$\}$,'' in \emph{16th USENIX Symposium on Operating Systems Design and Implementation (OSDI 22)}, 2022, pp. 303--320.

\bibitem{narayanan2021solving}
D.~Narayanan, F.~Kazhamiaka, F.~Abuzaid, P.~Kraft, A.~Agrawal, S.~Kandula, S.~Boyd, and M.~Zaharia, ``Solving large-scale granular resource allocation problems efficiently with pop,'' in \emph{Proceedings of the ACM SIGOPS 28th Symposium on Operating Systems Principles}, 2021, pp. 521--537.

\bibitem{ousterhout2013sparrow}
K.~Ousterhout, P.~Wendell, M.~Zaharia, and I.~Stoica, ``{Sparrow: Distributed, Low Latency Scheduling},'' in \emph{SOSP}, 2013.

\bibitem{park20183sigma}
J.~W. Park, A.~Tumanov, A.~Jiang, M.~A. Kozuch, and G.~R. Ganger, ``3sigma: distribution-based cluster scheduling for runtime uncertainty,'' in \emph{EuroSys}, 2018.

\bibitem{qiu2020firm}
H.~Qiu, S.~S. Banerjee, S.~Jha, Z.~T. Kalbarczyk, and R.~K. Iyer, ``Firm: An intelligent fine-grained resource management framework for slo-oriented microservices,'' \emph{arXiv preprint arXiv:2008.08509}, 2020.

\bibitem{romero2021faast}
F.~Romero, G.~I. Chaudhry, {\'I}.~Goiri, P.~Gopa, P.~Batum, N.~J. Yadwadkar, R.~Fonseca, C.~Kozyrakis, and R.~Bianchini, ``{Faa\$T: A Transparent Auto-Scaling Cache for Serverless Applications},'' \emph{arXiv preprint arXiv:2104.13869}, 2021.

\bibitem{rzadca2020autopilot}
K.~Rzadca, P.~Findeisen, J.~Swiderski, P.~Zych, P.~Broniek, J.~Kusmierek, P.~Nowak, B.~Strack, P.~Witusowski, S.~Hand, and J.~Wilkes, ``Autopilot: workload autoscaling at google,'' in \emph{Proceedings of the Fifteenth European Conference on Computer Systems}, 2020, pp. 1--16.

\bibitem{shahrad2020serverless}
M.~Shahrad, R.~Fonseca, {\'I}.~Goiri, G.~Chaudhry, P.~Batum, J.~Cooke, E.~Laureano, C.~Tresness, M.~Russinovich, and R.~Bianchini, ``Serverless in the wild: Characterizing and optimizing the serverless workload at a large cloud provider,'' in \emph{2020 USENIX Annual Technical Conference (USENIX ATC 20)}, 2020, pp. 205--218.

\bibitem{sriraman2018mu}
A.~Sriraman and T.~F. Wenisch, ``$\mu$ suite: a benchmark suite for microservices,'' in \emph{2018 IEEE International Symposium on Workload Characterization (IISWC)}.\hskip 1em plus 0.5em minus 0.4em\relax IEEE, 2018, pp. 1--12.

\bibitem{Sriraman18}
A.~Sriraman and T.~F. Wenisch, ``{\textmu}tune: Auto-tuned threading for {OLDI} microservices,'' in \emph{13th {USENIX} Symposium on Operating Systems Design and Implementation ({OSDI} 18)}.\hskip 1em plus 0.5em minus 0.4em\relax Carlsbad, CA: {USENIX} Association, Oct. 2018, pp. 177--194.

\bibitem{suresh2017distributed}
L.~Suresh, P.~Bodik, I.~Menache, M.~Canini, and F.~Ciucu, ``Distributed resource management across process boundaries,'' in \emph{Proceedings of the 2017 Symposium on Cloud Computing}.\hskip 1em plus 0.5em minus 0.4em\relax ACM, 2017, pp. 611--623.

\bibitem{netflix_decompose}
{Tony Mauro}, ``{Adopting microservices at Netflix: Lessons for architectural design},'' \url{https://www.nginx.com/blog/microservicesat-netflix-architectural-best-practices/}.

\bibitem{tumanov2016tetrisched}
A.~Tumanov, T.~Zhu, J.~W. Park, M.~A. Kozuch, M.~Harchol-Balter, and G.~R. Ganger, ``{TetriSched: global rescheduling with adaptive plan-ahead in dynamic heterogeneous clusters},'' in \emph{EuroSys}, 2016.

\bibitem{vaswani2017attention}
A.~Vaswani, N.~Shazeer, N.~Parmar, J.~Uszkoreit, L.~Jones, A.~N. Gomez, {\L}.~Kaiser, and I.~Polosukhin, ``Attention is all you need,'' \emph{Advances in neural information processing systems}, vol.~30, 2017.

\bibitem{wang2022deepscaling}
Z.~Wang, S.~Zhu, J.~Li, W.~Jiang, K.~Ramakrishnan, Y.~Zheng, M.~Yan, X.~Zhang, and A.~X. Liu, ``Deepscaling: microservices autoscaling for stable cpu utilization in large scale cloud systems,'' in \emph{Proceedings of the 13th Symposium on Cloud Computing}, 2022, pp. 16--30.

\bibitem{welch1947generalization}
B.~L. Welch, ``The generalization of ‘student's’problem when several different population varlances are involved,'' \emph{Biometrika}, vol.~34, no. 1-2, pp. 28--35, 1947.

\bibitem{welsh2001seda}
M.~Welsh, D.~Culler, and E.~Brewer, ``Seda: An architecture for well-conditioned, scalable internet services,'' \emph{ACM SIGOPS operating systems review}, vol.~35, no.~5, pp. 230--243, 2001.

\bibitem{powerchief}
H.~Yang, Q.~Chen, M.~Riaz, Z.~Luan, L.~Tang, and J.~Mars, ``Powerchief: Intelligent power allocation for multi-stage applications to improve responsiveness on power constrained cmp,'' in \emph{Proceedings of the 44th Annual International Symposium on Computer Architecture}, ser. ISCA ’17.\hskip 1em plus 0.5em minus 0.4em\relax New York, NY, USA: Association for Computing Machinery, 2017, p. 133–146.

\bibitem{zhang2021sinan}
Y.~Zhang, W.~Hua, Z.~Zhou, G.~E. Suh, and C.~Delimitrou, ``Sinan: Ml-based and qos-aware resource management for cloud microservices,'' in \emph{Proceedings of the 26th ACM International Conference on Architectural Support for Programming Languages and Operating Systems}, 2021, pp. 167--181.

\bibitem{zhou2018overload}
H.~Zhou, M.~Chen, Q.~Lin, Y.~Wang, X.~She, S.~Liu, R.~Gu, B.~C. Ooi, and J.~Yang, ``Overload control for scaling wechat microservices,'' in \emph{Proceedings of the ACM Symposium on Cloud Computing}.\hskip 1em plus 0.5em minus 0.4em\relax ACM, 2018, pp. 149--161.

\bibitem{zhou2018fault}
X.~Zhou, X.~Peng, T.~Xie, J.~Sun, C.~Ji, W.~Li, and D.~Ding, ``Fault analysis and debugging of microservice systems: Industrial survey, benchmark system, and empirical study,'' \emph{IEEE Transactions on Software Engineering}, vol.~47, no.~2, pp. 243--260, 2018.

\end{thebibliography}

\end{document}